\documentclass[letter,11pt]{article}

\pdfoutput=1 

\usepackage{jheppub} 

\usepackage[sort&compress]{natbib} 
\usepackage{amsmath}
\usepackage{graphicx}
\usepackage{dcolumn}
\usepackage{bm}
\usepackage{bbm}
\usepackage{epsfig}
\usepackage{slashed}
\usepackage{amssymb}
\usepackage{color}
\usepackage[font=small]{caption}
\usepackage[font=small]{subcaption}
\usepackage{tabulary}
\usepackage{soul}

\makeatletter

\newcommand{\newc}{\newcommand}
\newc{\renewc}{\renewcommand}

\def\beq{\begin{equation}}
\def\eeq{\end{equation}}
\def\bea{\begin{eqnarray}}
\def\eea{\end{eqnarray}}
\def\bitem{\begin{itemize}}
\def\eitem{\end{itemize}}
\def\ba{\begin{array}}
\def\ea{\end{array}}
\def\bal{\begin{align}}
\def\eal{\end{align}}
\def\bi{\begin{itemize}}
\def\ei{\end{itemize}}
\def\lsim{\mathrel{\rlap{\lower4pt\hbox{\hskip1pt$\sim$}}
    \raise1pt\hbox{$<$}}}         
\def\gsim{\mathrel{\rlap{\lower4pt\hbox{\hskip1pt$\sim$}}
    \raise1pt\hbox{$>$}}}
\newc{\red}{\textcolor{red}}
\newc{\blue}{\textcolor{blue}}

\newc{\ie}{{\it i.e.~}}          \newc{\etal}{{\it et al.~}}
\newc{\eg}{{\it e.g.~}}          \newc{\etc}{{\it etc.~}}
\newc{\cf}{{\it c.f.~}}
\newc{\vs}{{\it vs.~}}
\newc{\os}{\mbox{\hspace{4pt}}}
\newc{\us}{\mbox{\hspace{12pt}}}

\renewc{\bar}{\overline}

\newc{\gev}{\,{\rm GeV}}
\newc{\mev}{\,{\rm MeV}}
\newc{\ev}{\,{\rm eV}}
\newc{\kev}{\,{\rm keV}}
\newc{\tev}{\,{\rm TeV}}

\newc{\LM}{\mathcal{L}}
\newc{\SM}{\mathcal{S}}

\newc{\HM}{\mathcal{H}}
\newc{\GM}{\mathcal{G}}
\newc{\OM}{\mathcal{O}}
\newc{\FM}{\mathcal{F}}
\newc{\AM}{\mathcal{A}}
\newc{\BM}{\mathcal{B}}
\newc{\NM}{\mathcal{N}}
\newc{\WM}{\mathcal{W}}
\newc{\ZM}{\mathcal{Z}}

\newc{\Chi}{\mathcal{X}}

\definecolor{red1}{cmyk}{0,1,1,0.1}
\definecolor{blue1}{cmyk}{1,0,0,0}

\title{Electroweak Kaluza-Klein  Dark Matter}

\author[a]{Thomas Flacke,}
\author[a,b]{Dong Woo Kang,}
\author[c,d]{Kyoungchul Kong,}
\author[c,e]{Gopolang Mohlabeng,}
\author[f,g]{Seong Chan Park}

\affiliation[a]{Center for Theoretical Physics of the Universe,
Institute for Basic Science (IBS), \\Daejeon 34051, Korea}
\affiliation[b]{Department of Physics, Sungkyunkwan University, Suwon 16419, Korea}
\affiliation[c]{Department of Physics and Astronomy, University of Kansas, Lawrence, KS 66045, USA}
\affiliation[d]{Pittsburgh Particle physics, Astrophysics, and Cosmology Center, Department of Physics and Astronomy, University of Pittsburgh, Pittsburgh, PA 15260, USA.}
\affiliation[e]{Fermi National Accelerator Laboratory, Theory Group, Batavia, IL}
\affiliation[f]{Department of Physics and IPAP, Yonsei University, Seoul 03722, Korea }
\affiliation[g]{Korea Institute for Advanced Study, Seoul 02455, Korea}

\emailAdd{flacke@ibs.re.kr}
\emailAdd{kdwgod54@skku.edu}
\emailAdd{kckong@ku.edu}
\emailAdd{gopolang.mohlabeng@ku.edu}
\emailAdd{sc.park@yonsei.ac.kr}

\abstract{
In models with universal extra dimensions (UED), the lightest Kaluza-Klein excitation of neutral electroweak gauge bosons is a stable, weakly interacting massive particle and thus is a candidate for dark matter thanks to Kaluza-Klein parity. 
We examine concrete model realizations of such dark matter in the context of non-minimal UED extensions.  
The boundary localized kinetic terms for the electroweak gauge bosons lead to a non-trivial mixing among the first Kaluza-Klein excitations of the ${\rm SU}(2)_W$ and ${\rm U}(1)_Y$ gauge bosons and the resultant low energy phenomenology is rich. 
We investigate implications of various experiments including low energy electroweak precision measurements, direct and indirect detection of dark matter particles and direct collider searches at the LHC. Notably, we show that the electroweak Kaluza-Klein dark matter can be as heavy as 2.4 TeV, which is significantly higher than $1.3$ TeV as is indicated as an upper bound in the minimal UED model. \\}


\begin{document} 

\hspace*{129mm}{\large \tt CTPU-16-31} \\
\hspace*{112mm}{\large \tt FERMILAB-PUB-16-554-T} \\
\hspace*{127mm}{\large \tt PITT-PACC-1616}

\maketitle
\flushbottom


\section{Introduction}
\label{sec:intro}

 Arguably the most attractive dark matter (DM) candidate in particle physics would be a weakly interacting massive particle (WIMP) thanks to the ``WIMP miracle".  The relic abundance of WIMPs at the weak energy scale lies in the right range to explain the observed dark matter abundance in the Universe: with the thermal averaged self-annihilation cross section in a typical range,  $\langle \sigma v \rangle \simeq 1$ pb, the observed amount of DM density is naturally accounted as $$\Omega_{\rm DM} h^2 \simeq \frac{0.1 {\rm pb}}{{\langle \sigma v \rangle}} \simeq 0.1,$$ where $\Omega_{DM} = \rho_{DM}/\rho_c$ is the relic density of the WIMP in the unit of the critical density, $\rho_c = 3H_0^2/8\pi G$, with the Hubble constant $H_0 = 100 \, h \, {\rm km\, s^{-1} Mpc^{-1}}$ and the gravitational constant $G \simeq 6.67 \times 10^{-11} {\rm m^3 kg^{-1} s^{-1}} $. The recent measurement by Planck gives $h\simeq 0.68$~\cite{Ade:2013zuv} \footnote{The local value of the Hubble constant recently updated by observations of Cepheid variables is a bit higher, $h \simeq 0.73$ \cite{Riess:2016jrr}, which may be understood with new relativistic degrees of freedom \cite{Park:2013bza}.}. A disadvantageous aspect of the WIMP miracle, however, is that one cannot learn the detailed nature of the WIMP solely from the DM amount  because WIMP candidates in a wide mass range with different spins and different interaction patterns still provide essentially the
same predictions.  To overcome this ``degeneracy problem'', one has to rely on complementary approaches by measuring various properties of DM relying on direct and indirect detection experiments as well as collider experiments.  A promising DM candidate is found in the minimal supersymmetric extension of the SM (MSSM) where the lightest supersymmetric particle (neutralino in a wide range of model parameters) is a WIMP candidate. In general, a neutralino is composed of electroweakinos and also higgsinos~\cite{Jungman:1995df}. The stability of the neutralino is guaranteed by R-parity. Not so surprisingly,  when the mass of the neutralino is assumed to be at the weak scale, as is required to address the hierarchy problem, the corresponding thermal relic abundance is in the right range due to the WIMP miracle. However, the full properties of DM sensitively depends on the detailed composition of the neutralino. Depending on the dominant neutralino component DM is Higgsino-like, photino-like or Zino ($W_3$-ino)-like. Generically, all components can be present, making SUSY DM phenomenology very rich. 

In this paper we examine an alternative attractive candidate, Kaluza-Klein (KK) DM, in flat extra dimensions. A minimal KK DM has been discussed in universal extra dimension (UED) models \cite{Appelquist:2000nn} based on a TeV scale extra dimension \cite{Antoniadis:1990ew} where 
{ the entire standard model (SM) particle content is assumed to be realized as the zero KK modes of scalar, fermion, and gauge fields (with the SM field quantum numbers) which propagate on a space-time $\mathcal{M}_4\times X$, where $\mathcal{M}_4$ is 4-dimensional Minkowski space and $X$ is a compact flat space of extra dimension(s).} 
{ In 4+1 dimensions, chiral zero mode fermions can be obtained when the extra dimensional space is taken to be the orbifold $S^1/\mathbb{Z}_2$ (or equivalently the interval $\left[-L, L\right]$ where $L=\pi R/2$ with the compactification radius $R$) which we focus on in this article.}\footnote{Fermions on $\mathcal{M}_4\times S_1$ are vectorlike. However by orbifolding, half of the spinor degrees of freedom are projected out due to the boundary conditions imposed at the orbifold fixed points which results in a chiral zero mode for each fermion field after KK decomposition.} In UED, all the SM fields are accompanied by their KK excitations with a mass gap of the order of the inverse compactification radius $1/R$. One of the attractive features of UED is KK parity conservation. KK parity is the reflection symmetry about the mid point of the extra dimension. It represents a geometric $\mathbb{Z}_2$ symmetry which is stable against quantum corrections and is thereby conserved if imposed at tree level ~\cite{Georgi:2000wb,Cheng:2002iz}. KK parity protects the lightest KK particle (LKP) from decay~\cite{Servant:2002aq, Cheng:2002ej, Cheng:2002ab,Datta:2010us,Datta:2005zs}.  In minimal UED (MUED) \cite{Cheng:2002iz}, the first KK excitation of the photon\footnote{Actually the LKP photon~($\gamma^{(1)}$) is very close to the KK excitation of the hypercharge gauge boson~($B^{(1)}$), because the weak mixing angle for the KK states are suppressed by a small factor $(m_W^2/m_{KK}^2)\ll 1$ \cite{Cheng:2002iz}.} with a mass $M_1=1/R$ is  the LKP.\footnote{See Refs.\cite{Hooper:2007qk,Servant:2014lqa} for reviews on universal extra dimensions as well as Refs. \cite{Aad:2015mia,Deutschmann:2017bth,Beuria:2017jez} for the most recent LHC bounds on MUED.}
The phenomenology of KK DM \cite{Bertone:2010ww,Kong:2005hn,Burnell:2005hm,Kakizaki:2006dz,Belanger:2010yx} becomes much richer when bulk mass terms for fermions \cite{Park:2009cs, Chen:2009gz,Kong:2010qd,Csaki:2010az,Kim:2011tq,Flacke:2011nb} and the boundary localized kinetic terms (BLKTs) \cite{Dvali:2001gm, Carena:2002me, delAguila:2006atw} are allowed as in non-minimal UED (NMUED)~\cite{Gao:2014wga,Flacke:2008ne,Datta:2012tv,Dey:2013cqa, Flacke:2013pla, Flacke:2012ke,Flacke:2013nta,Flacke:2014jwa,Ishigure:2016kxp}.  We note that the boundary localized terms and the bulk mass terms are compatible with the Lorentz symmetry and the gauge symmetries of the model so that such terms should be included in the generic effective field theory action~\cite{Flacke:2008ne}.

 The presence of BLKTs for electroweak gauge bosons modifies the composition of the LKP, which appears as a mixture of KK excitations of the hyper-charge gauge boson, $B^{(1)}$ and the neutral component of the weak gauge boson $W^{(1)}_3$~\cite{Flacke:2008ne}. 
 This is  different from MSSM, since no KK Higgs component is involved due to different spin of the KK partners of the KK Higgs as compared to KK gauge bosons. Several studies have considered the KK photon and the KK Z boson LKP separately as DM candidates~\cite{Arrenberg:2008wy,Arrenberg:2013paa,Flacke:2009eu,Datta:2013nua}. Here, we  consider generic mixing in electroweak KK DM sector and study various phenomenological aspects of KK DM in a more general framework of NMUED \footnote{In this article we focus on DM in 5D models compactified on $S_1/\mathbb{Z}_2$. For DM in different compactifications and its phenomenology, see Refs. \cite{Dobrescu:2004zi,Dobrescu:2007ec,Maru:2009wu,Cacciapaglia:2009pa,Arbey:2012ke,Cacciapaglia:2016xty,Andriot:2016rdd,Freitas:2007rh,Dobrescu:2007xf}.}.
In MUED all the BLKTs are chosen to vanish at the cutoff scale and quantities at electroweak scale are obtained by renormalization group equations. In this article, we take the BLKTs as free parameters at the compactification scale instead. As a result, mixings and mass spectra are modified as compared to MUED.

This article is structured as follows: In section \ref{sec:setup0}, we present the model of electroweak boson KK DM allowing BLKTs in NMUED and examine KK spectra and mixings among KK states. In section \ref{sec:bounds}, we discuss current collider and precision measurement bounds on the given setup focusing on the effects of allowed four-Fermi operators, as well as collider constraints from the LHC. In section \ref{sec:DM}, we study the impact of BLKTs on the relic abundance of electroweak KK DM and on the direct detection rates taking the latest bounds into account. Finally, we conclude in section \ref{sec:conclusion}.

\section{The Setup}
\label{sec:setup0}

In this section we set up the model Lagrangian and discuss the KK decomposition of the electroweak KK bosons in the presence of  BLKTs. We focus on mixings among KK weak gauge bosons.  We will follow notations as in a recent review, Ref.~\cite{Flacke:2013pla}.

\subsection{Model Lagrangian} \label{sec:setup}

When we embed the SM in a five dimensional space $\mathcal{M}_4\times[-L,L]$, the UED action is given in the following form:
\beq
S_5=\int d^4x \int_{-L}^L dy  \, \left[ {\cal L}_V+{\cal L}_{\Psi}+{\cal L}_H+{\cal L}_{\rm Yuk}\right] ,
\label{5Daction}
\eeq
where  $y=\pm L$ are the orbifold fixed points, which are the boundaries of the fifth dimension.  The kinetic energy of the gauge bosons and fermions propagating in $5D$ bulk are ${\cal L}_V$ and ${\cal L}_\Psi$. The Lagrangian for the Higgs boson and the Yukawa interactions with fermions are ${\cal L}_H$ and ${\cal L}_{\rm Yuk}$, respectively. 
The explicit form of each term is given as follows:
\bea
{\cal L}_V&=& \sum_{\AM}^{G,W,B} -\frac{1}{4} \AM^{MN}\cdot \AM_{MN}\,, \label{Vlag}\\
{\cal L}_{\Psi}&=& \sum_{\Psi}^{Q,U,D,L,E}i \overline{\Psi} {D}_M \Gamma^M \Psi\,, \label{FLag}\\
{\cal L}_{H}&=&\left(D_\mu H\right)^\dagger D^\mu H+ \mu_5^2 |H|^2- \lambda_5 |H|^4\,,\label{HiggsLag}  \\
{\cal L}_{\rm Yuk}&=&\lambda_5^E\overline{L}HE+ \lambda_5^D\overline{Q}HD+\lambda_5^U\overline{Q}\tilde{H}D+\mbox{h.c.}\,, \label{eq:boundaryyukawa}
\eea
where $\AM$ denotes the five dimensional gauge bosons in the SM gauge group, i.e., the gluon ($G$), weak gauge bosons ($W$) and the hypercharge gauge boson ($B$). $D_M = \partial_M +i \hat{g}_{3} \lambda\cdot G_{M}+  i \hat{g}_{2}  \tau \cdot W_{M}  +  i\hat{g}_{1} Y B_{M}$ is the gauge covariant derivatives, where the $\hat{g}_{i}$'s are the five dimensional couplings of the SM, and $\lambda$'s and $\tau$'s are the generators of ${\rm SU(3)_c}$ and ${\rm SU(2)_W}$, respectively.
The fermions, $\Psi = L, E,Q,D,U$ are Dirac spinors containing both chiralities in the KK decomposition as $\Psi (x,y) = \sum_n \psi^n_L(x) f^n_L(y) +\psi^n_R(x) f^n_R(y)$ where $\psi^n_{L/R}(x)$ is the $n$-th KK excitation mode with left-(right-) chirality, respectively and $f^n_{L/R}(y)$ is the corresponding KK basis function in the fifth dimension. The model is 5D Lorentz symmetric and the SM gauge symmetries are assumed as the internal symmetries. One should notice that the constructed Lagrangian is invariant under the inversion ($y \to -y$), such that the model respects the Kaluza-Klein parity (KK-parity). 
From the kinetic terms one can read out the mass dimensions of the fields and the coupling constants: $[\AM] = [H] = {\rm Mass}^{3/2}$, $[\Psi]={\rm Mass^2}$, $[\mu_5]={\rm Mass}$, $[\lambda_5^\Psi]={\rm Mass}^{-1/2}$ and $[\hat{g}_i]={\rm Mass}^{-1/2}$.  The KK basis functions are dimensionful as $[f_{L/R}^n]={\rm Mass}^{1/2}$ and the KK modes are regarded as the conventional fields in 4D, $[\psi_{L/R}^n]={\rm Mass}^{3/2}$.

Notably, the 4D spacetime symmetry and the gauge symmetries of the model allow additional boundary localized operators. {Even if the absence of such operators is assumed at tree level, they are induced by radiative corrections \cite{Georgi:2000wb,Cheng:2002iz}, which shows that these operators cannot be forbidden by an underlying symmetry and their coefficients should thus be considered as additional parameters of the model which can only be calculated from the (so far unknown) UV completion of the model. If the UV completion respects KK-parity, the boundary terms on the two orbifold fixed points are related.\footnote{Apart from KK parity conserving boundary terms, UED models can also contain KK parity odd fermion masses in the bulk whilst preserving KK parity in all interactions~\cite{Park:2009cs}. For studies of UED models with KK parity violating boundary terms {\it c.f. e.g.} \cite{Datta:2012xy,Datta:2013lja,Shaw:2014gba}.} }  
 In this article, we focus on the boundary localized terms for the electroweak gauge bosons respecting the KK-parity, the lightest combination of which would serve as dark matter:
\beq
S_{bdy}= \int d^4 x \int_{-L}^L d y  \, \left(-\frac{r_W}{4}   W_{\mu\nu} \cdot W^{\mu\nu}-\frac{r_B}{4}   B_{\mu\nu} B^{\mu\nu},  \right) \left[\delta(y-L)+\delta(y+L)\right],
\eeq
where $r_W$ and $r_B$ are parameters describing the strength of the boundary localized terms and their mass dimensions are $[r_W]=[r_B]={\rm Mass}^{-1}$. 

The boundary localized operators for the Higgs would affect the electroweak symmetry breaking in general but the KK state of the Higgs boson would not mix with electroweak gauge bosons because of the different spins. This makes a clear distinction from the MSSM where a neutralino is a mixture of higgsinos and electroweakinos.

The boundary localized terms modify the KK mass spectra and the KK wave functions of the electroweak gauge bosons, as will be worked out in detail in the next section. This in turn has important implications for the dark matter phenomenology: {\it (i)} Due to the modified masses of the electroweak gauge bosons at the first KK level, the UED dark matter candidate now becomes a linear combination of the $B^{(1)}$ and the $W^{3(1)}$ with the mixing angles determined by $r_W, r_B$, and $R^{-1}$, {\it (ii)} due to the modified wave functions, the couplings amongst the electroweak gauge bosons and the fermions (which follow from the overlap integrals of wave functions) are modified. Therefore 
the parameter space $(r_W,r_B,R^{-1})$ will be constrained by various tests such as electroweak precision measurement and collider searches.

\bigskip

 In this article, we only focus on the boundary terms for electroweak gauge bosons. Therefore, our results by no means cover the entire NMUED parameter space, but rather show the main effects of changing the LKP from a $B^{(1)}$ to a $W^{3(1)}$ dark matter candidate, and its correlated implications for collider searches and precision bounds.

\subsection{Kaluza Klein decomposition}\label{sec:KKdecomp}

KK masses and wave functions for the KK fermions, the KK gluon and the KK Higgs are given by the standard UED results (no boundary terms for these)
\begin{eqnarray}
f^e(y)=\left\{\begin{matrix}
f^e_0&=&\sqrt{\frac{1}{2L}} \, ,\\ 
f^e_{2n}&=&\sqrt{\frac{1}{L}}\cos \frac{2ny}{R} \, ,\\ 
f^e_{2n+1}&=&\sqrt{\frac{1}{L}}\sin \frac{(2n+1)y}{R}
\end{matrix}\right. \, ,\\
f^o(y)=\left\{\begin{matrix}
f^o_{2n+1}&=&\sqrt{\frac{1}{L}}\cos \frac{(2n+1)y}{R} \, ,\\ 
f^o_{2n}&=&\sqrt{\frac{1}{L}}\sin\frac{2ny}{R} \, ,
\end{matrix}\right.
\end{eqnarray}
where $f^e$ denote the KK wave functions of the $\mathbb{Z}_2$ even fields $G_\mu, Q_L,U_R,D_R,L_L,E_R, h$, and $f^o$ denote the KK wave functions of the $\mathbb{Z}_2$ odd fields $G_5,Q_R,U_L,D_L,L_R,E_L$.
The wave functions satisfy the normalization condition $\int_{-L}^L dy f_n^* f_m =\delta_{mn}$, and the masses are determined by $m^2_{\Phi^{(n)}}=(n/R)^2+m^2_{\Phi^{(0)}}$, with the zero mode mass  $m^2_{\Phi^{(0)}}$, given by the Higgs mechanism. 
Note that $[f^{e/o}]={\rm Mass}^{1/2} ={\rm Length}^{-1/2}$, which is consistent with the Kronecker-delta normalization for orthonormal basis.

For the electroweak gauge bosons
the boundary kinetic terms modify the wave functions.
The KK decomposition of electroweak gauge bosons in the presence of boundary kinetic terms have been performed in Ref.~\cite{Flacke:2008ne}. Treating electroweak symmetry breaking as a perturbation, the gauge fields are decomposed as
\bea
W _\mu(x,y)&=&\sum_{n=0}^\infty W_\mu^{(n)}(x)f^W_n(y) \,,\\
B _\mu(x,y)&=&\sum_{n=0}^\infty B_\mu^{(n)}(x)f^B_n(y)\,,
\eea
where
\bea
f^{W/B}_n(y)=\begin{cases}
\mathcal{N}^{W/B}_0  & \text{ if } n=0, \\ 
\mathcal{N}^{W/B}_n  \sin(k^{W/B}_n y) & \text{ if } n=odd, \\ 
\mathcal{N}^{W/B}_n \cos( k^{W/B}_n y) & \text{ if } n=even, 
\end{cases}
\eea
with the normalization factors
\bea
\mathcal{N}^{W/B}_{n}= \begin{cases}
 \frac{1}{ \sqrt{ 2 L ( 1 + \frac{r_{W/B}}{L})}} & \text{ if } n=0,\\
 \frac{1}{\sqrt{L+ r_{W/B} \sin^2 (k^{W/B}_n L)}}  & \text{ if } n=odd, \\ 
 \frac{1}{\sqrt{L+ r_{W/B} \cos^2 (k^{W/B}_n L)} } & \text{ if } n=even.
\end{cases}
\eea
The wave numbers $k_n$ are determined by 
\bea
 \cot (k^{W/B}_n L) &=  r_{W/B} k^{W/B}_n  \,    & \text{ if } n=odd,  \\\nonumber
 \tan (k^{W/B}_n L) &= - r_{W/B} k^{W/B}_n  \,   &\text{ if } n=even \, .
 \label{gaugeKKmass}
\eea
Furthermore the wave functions satisfy the orthogonality relations
\beq
\int_{-L}^L dy f^{W/B}_mf^{W/B}_n\left[1+r_{W/B}\left(\delta(y+L)+\delta(y-L)\right)\right]=\delta_{mn}.
\label{gaugesclprd}
\eeq
Again $[f^{W/B}_n]={\rm Mass}^{1/2}={\rm Length}^{-1/2}$,  which is consistent with our normalization conditions.

Finally, the effective 4D action of the electroweak gauge bosons is obtained after integrating over $y$:
\bea
S_{4D}\ni \int d^{4}x &&\left\{\sum_n \left[-\frac{1}{4} \sum_n B^{(n)\mu\nu}B^{(n)}_{\mu\nu}-\frac{\left(k^{B}_n\right)^2}{2} B^{(n)\mu}B^{(n)}_\mu\right.\right.\nonumber\\
&& \hspace{30pt}\left.\left.-\frac{1}{4} \sum_n W^{(n)a\mu\nu}\cdot W^{(n)a}_{\mu\nu}-\frac{\left(k^{W}_n\right)^2}{2} W^{(n)a\mu}W^{(n)a}_\mu\right]\right.\\
&&\hspace{10pt}\left.  +\sum_{m,n} \left[-\frac{\hat{g}^2_1  v^2}{8} \mathcal{F}^{BB}_{mn}B^{(m)\mu}B^{(n)}_\mu-\frac{\hat{g}_1\hat{g}_2  v^2}{8} \mathcal{F}^{WB}_{mn}B^{(m)\mu}W^{(n)3}_\mu\right.\right.\nonumber\\
&&\hspace{45pt}\left.\left.-\frac{\hat{g}^2_2  v^2}{8} \mathcal{F}^{WW}_{mn}W^{(m)a\mu}W^{(n)a}_\mu\right]\right\}\,,\nonumber
\eea
where $\hat{g}_{1,2}$ and $ v$ are the 5D ${\rm U(1)}_Y$ and ${\rm SU(2)}_W$ gauge couplings and the vacuum expectation value. 
The mixing parameters are defined as 
\bea
\mathcal{F}^{BB}_{mn}&=&\int_{-L}^{L} \, \frac{dy}{2L}\, f^{B}_m(y)f^{B}_n(y) \, ,\nonumber\\
\mathcal{F}^{BW}_{mn}&=&\int_{-L}^{L} \,  \frac{dy}{2L}\, f^{B}_m(y)f^{W}_n(y)\, ,\nonumber\\
\mathcal{F}^{WW}_{mn}&=&\int_{-L}^{L} \,  \frac{dy}{2L}\, f^{W}_m(y)f^{W}_n(y)\,,
\eea
where the  normalization factor, $1/(2L)$, comes from the normalization factor of the zero mode Higgs vacuum expectation value. The resultant mass dimensions of the mixing parameters are $[\mathcal{F}_{mn}^{VV'}]={\rm Mass}$ for $V(V') = B$ or $W$. It should be noted that $\mathcal{F}^{BB}_{mn}$ and $\mathcal{F}^{WW}_{mn}$ are not orthogonal in our basis as they are orthogonal with respect to the scalar product as in Eq.~(\ref{gaugesclprd}) which includes the boundary parameters. The electroweak symmetry breaking terms having $v^2$ thus induce KK-mode-mixing in the basis we are using. 
Note that KK parity is still conserved so that even and odd modes do not mix.

We can separately analyze the mass matrices for  KK even modes and odd modes. The matrix for even modes is relevant for tree level modifications of zero mode couplings as well as the couplings of the zero modes and the second (and higher even) KK modes.  These are particularly important in $Z'$-like new gauge boson searches since the production and decay of $Z'=Z_2$ would be decided by the matrix. It is also important to consider 4-fermion operators among zero mode fermions, which are induced by even KK mode exchange. They can be probed by precision measurements, which will be analyzed in section \ref{sec:bounds}. The mass matrix for KK odd modes is particularly relevant for the DM physics since the nature and the structure of the couplings of the LKP (the lightest odd mode) is determined by the mass matrix.

\subsection{Mass matrices and mixing angles of KK gauge bosons}

The mass matrix of the even-numbered neutral mass matrix in the $B^{(2n)}$ - $W^{3(2n)}$ basis reads 
\beq
M^2_{n,e} =\left(
\begin{array}{ccccc}
\frac{\hat{g_1}^2 v^2}{4}\mathcal{F}^{BB}_{00} & \frac{\hat{g_1}\hat{g_2} v^2}{4}\mathcal{F}^{BW}_{00} & 
\frac{\hat{g_1}^2 v^2}{4}\mathcal{F}^{BB}_{02} & \frac{\hat{g_1}\hat{g_2} v^2}{4}\mathcal{F}^{BW}_{02} & \hdots\\ 
\frac{\hat{g_1}\hat{g_2} v^2}{4}\mathcal{F}^{BW}_{00} & \frac{\hat{g_2}^2 v^2}{4}\mathcal{F}^{WW}_{00} &
\frac{\hat{g_1}\hat{g_2} v^2}{4}\mathcal{F}^{BW}_{02} & \frac{\hat{g_2}^2 v^2}{4}\mathcal{F}^{WW}_{02} & \hdots\\
\frac{\hat{g_1}^2 v^2}{4}\mathcal{F}^{BB}_{20} & \frac{\hat{g_1}\hat{g_2} v^2}{4}\mathcal{F}^{BW}_{20} & 
\left(k^B_2\right)^2+\frac{\hat{g_1}^2 v^2}{4}\mathcal{F}^{BB}_{22} & \frac{\hat{g_1}\hat{g_2} v^2}{4}\mathcal{F}^{BW}_{22} & \hdots\\ 
\frac{\hat{g_1}\hat{g_2} v^2}{4}\mathcal{F}^{BW}_{20} & \frac{\hat{g_2}^2 v^2}{4}\mathcal{F}^{WW}_{20} &
\frac{\hat{g_1}\hat{g_2} v^2}{4}\mathcal{F}^{BW}_{22} &\left(k^W_2\right)^2 + \frac{\hat{g_2}^2 v^2}{4}\mathcal{F}^{WW}_{22} & \hdots\\
\vdots&\vdots&\vdots&\vdots&\ddots
\end{array}\right).
\eeq

We can further simplify the mass matrix by using the fact that the zero mode wave functions are flat. First, let us define
\beq
g_{1,2}=\hat{g}_{1,2}\mathcal{N}^{B,W}_0,\label{g12def}
\eeq
and the ``normalized'' and dimensionless overlap integrals
\beq
\tilde{\mathcal{F}}^{BB}_{mn}\equiv \frac{\mathcal{F}^{BB}_{mn}}{(\mathcal{N}^{B}_0)^2} \,\,\, , \,\,\, \tilde{\mathcal{F}}^{WW}_{mn}\equiv \frac{\mathcal{F}^{WW}_{mn}}{(\mathcal{N}^{W}_0)^2} \,\,\, , \,\,\, \tilde{\mathcal{F}}^{BW}_{mn}\equiv \frac{\mathcal{F}^{BW}_{mn}}{\mathcal{N}^{B}_0\mathcal{N}^{W}_0}.\label{Fnorm}
\eeq
Then the mass matrix can be rewritten as 
\beq
M^2_{n,e} =\left(
\begin{array}{ccccc}
\frac{g_1^2 v^2}{4}& \frac{ g_1 g_2 v^2}{4} & 
\frac{ g_1^2 v^2}{4}\tilde{\mathcal{F}}^{BB}_{02} & \frac{ g_1 g_2 v^2}{4}\tilde{\mathcal{F}}^{BW}_{02} & \hdots\\ 
\frac{ g_1 g_2 v^2}{4} & \frac{ g_2^2 v^2}{4} &
\frac{ g_1 g_2 v^2}{4}\tilde{\mathcal{F}}^{BW}_{02} & \frac{ g_2^2 v^2}{4}\tilde{\mathcal{F}}^{WW}_{02} & \hdots\\
\frac{ g_1^2 v^2}{4}\tilde{\mathcal{F}}^{BB}_{20} & \frac{ g_1 g_2 v^2}{4}\tilde{\mathcal{F}}^{BW}_{20} & 
\left(k^B_2\right)^2+\frac{ g_1^2 v^2}{4}\tilde{\mathcal{F}}^{BB}_{22} & \frac{ g_1 g_2 v^2}{4}\tilde{\mathcal{F}}^{BW}_{22} & \hdots\\ 
\frac{ g_1 g_2 v^2}{4}\tilde{\mathcal{F}}^{BW}_{20} & \frac{ g_2^2 v^2}{4}\tilde{\mathcal{F}}^{WW}_{20} &
\frac{ g_1 g_2 v^2}{4}\tilde{\mathcal{F}}^{BW}_{22} &\left(k^W_2\right)^2 + \frac{ g_2^2 v^2}{4}\tilde{\mathcal{F}}^{WW}_{22} & \hdots\\
\vdots&\vdots&\vdots&\vdots&\ddots
\end{array}\right).
\eeq
Now, performing a field rotation on the zero modes
\beq
U^\dagger = \left(\begin{array}{ccc}
\cos\theta & \sin\theta & 0\\
-\sin\theta & \cos\theta & 0\\
0 & 0 & \mathbbm{1}
\end{array}
\right),
\eeq
with $\tan\theta = g_1/g_2$, one obtains
\begin{eqnarray}
U M^2_{n,e} U^\dagger &=&\left(
\begin{array}{ccccc}
0 & 0 & 0 & 0 & \hdots\\ 
0 & \frac{ (g^2_1 +g^2_2) v^2}{4} & \frac{  g_1 \sqrt{g_1^2+g^2_2} v^2}{4}\tilde{\mathcal{F}}^{BW}_{02} & \frac{ g_2 \sqrt{g_1^2+g^2_2} v^2}{4}\tilde{\mathcal{F}}^{WW}_{02} & \hdots\\
0 & \frac{  g_1 \sqrt{g_1^2+g^2_2} v^2}{4}\tilde{\mathcal{F}}^{BW}_{20} & \left(k^B_2\right)^2+\frac{ g_1^2 v^2}{4}\tilde{\mathcal{F}}^{BB}_{22} & \frac{ g_1 g_2 v^2}{4}\tilde{\mathcal{F}}^{BW}_{22} & \hdots\\ 
0 & \frac{ g_2 \sqrt{g_1^2+g^2_2} v^2}{4}\tilde{\mathcal{F}}^{WW}_{20} & \frac{ g_1 g_2 v^2}{4}\tilde{\mathcal{F}}^{BW}_{22} &\left(k^W_2\right)^2 + \frac{ g_2^2 v^2}{4}\tilde{\mathcal{F}}^{WW}_{22} & \hdots\\
\vdots&\vdots&\vdots&\vdots&\ddots
\end{array}\right)\, . \label{KKevenMass} 
\end{eqnarray}
In this basis, the masslessness of the photon is explicitly seen. As it is a linear combination of $B^{(0)}$ and $W^{(0)}_{3}$, both of which in this basis have flat wave functions, the photon wave function is also flat as expected for a massless particle.
At the same time we see that the zero mode of $Z$ (and $W$ as well)  mixes with the even KK modes of the $B$ and the $W_3$ in general. 

In our phenomenological study for dark matter physics, the most relevant mass matrix is the mass matrix for the first KK excitation of neutral gauge bosons. The lightest odd state would be the candidate of the DM:
\beq
\label{eq:odd}M^2_{n,odd}=\left(
\begin{array}{ccc}
\left(k^B_1\right)^2+\frac{\hat{g_1}^2\hat{v}^2}{4}\mathcal{F}^{BB}_{11} & \frac{\hat{g_1}\hat{g_2}\hat{v}^2}{4}\mathcal{F}^{BW}_{11} &  \hdots\\ 
\frac{\hat{g_1}\hat{g_2}\hat{v}^2}{4}\mathcal{F}^{BW}_{11} &\left(k^W_1\right)^2+ \frac{\hat{g_2}^2\hat{v}^2}{4}\mathcal{F}^{WW}_{11} & \hdots\\
\vdots&\vdots& \ddots
\end{array}\right).
\eeq
In the limit of vanishing boundary terms, $\mathcal{F}^{VV}_{nn}$ approaches the unity 
($\mathcal{F}^{VV}_{nn} \to 1$ for $r_W \to 0$ and $r_B \to 0$).
In addition to the terms from electroweak symmetry breaking, the boundary parameters play important roles here. They affect not only the overlap integrals $\mathcal{F}_{11}^{VV'}$ but also the value of the wave number $k_{1}^{B}$ and  $k_1^{W}$.

\begin{figure}
\centerline{
\includegraphics[width=.48\textwidth]{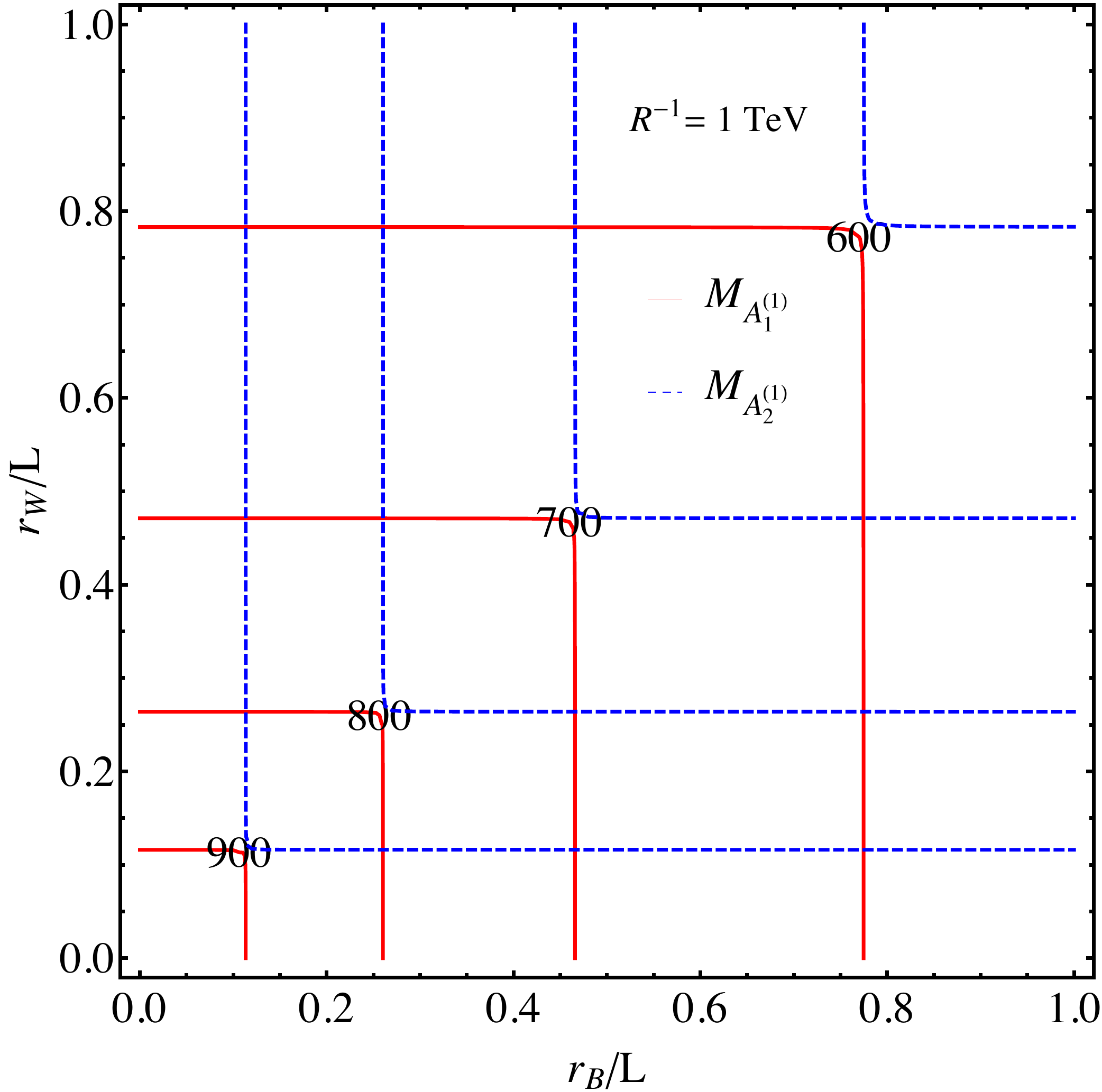} \hspace{0.1cm}
\includegraphics[width=.5\textwidth]{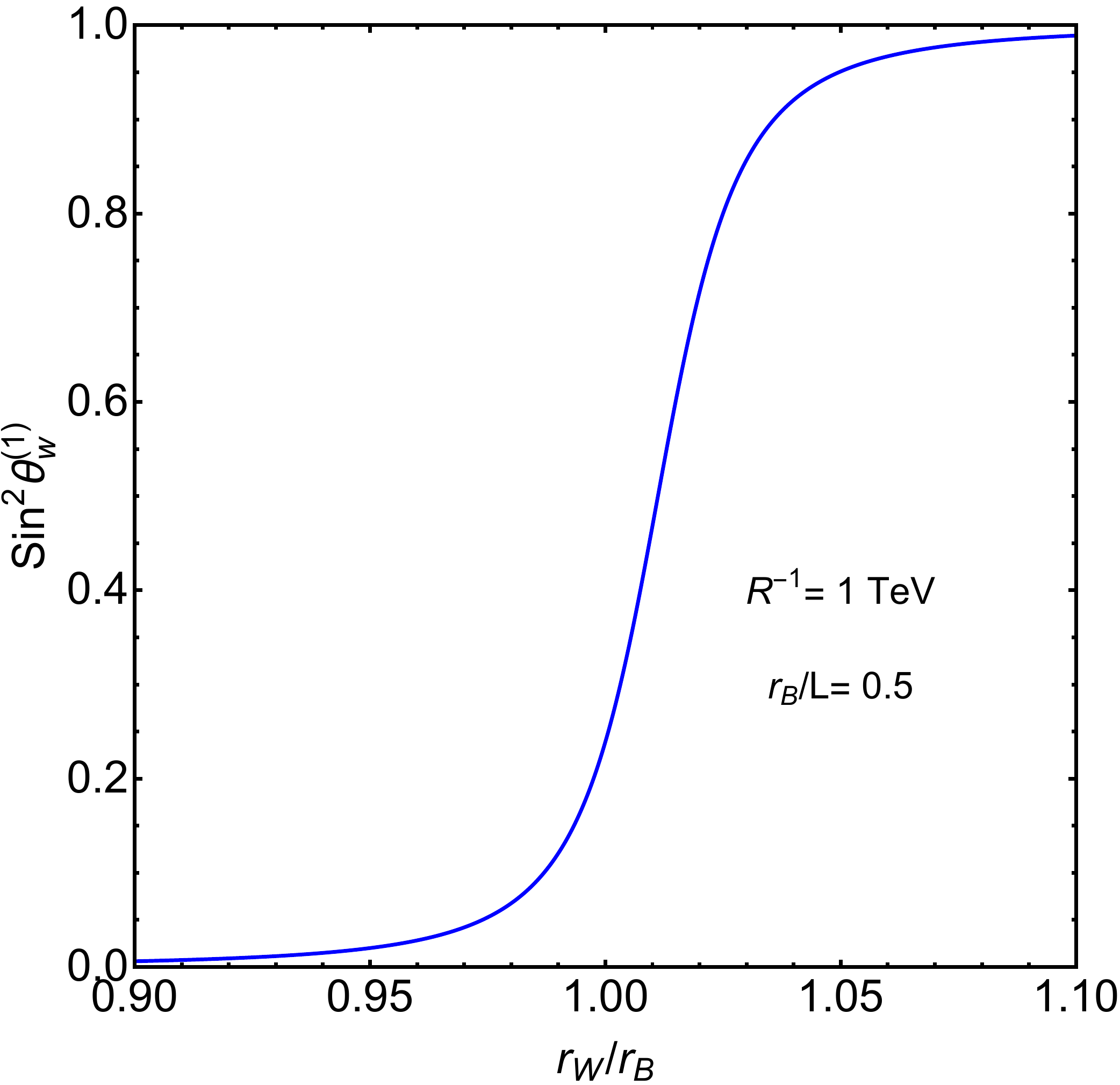}}
\caption{ Left: Contours of constant mass for the level 1 electroweak KK bosons $A_{1}^{(1)}$ and $A_{2}^{(1)}$. The contours were made assuming $R^{-1}$ = 1 TeV, and they show the mass dependence on the boundary terms $r_{B}$ and $r_{W}$. Right: The level 1 KK Weinberg angle $\sin^2\theta_{W}^{(1)}$ for $R^{-1}$ = 1 TeV and $r_{B}/L=0.5$.}
\label{A1A2contours}
\end{figure}

 The contours of the two lightest electroweak KK gauge boson masses is shown in the left panel of Fig.~\ref{A1A2contours}. We present the contour in the $(r_B, r_W)$ plane for $R^{-1}=1 \tev$ as an example. The red solid contour lines are for the lighter level one mass eigenstate $A_{1}^{(1)}$ while the blue dashed contours are for the heavier $A_{2}^{(1)}$ state. When a boundary parameter ($r_W$ or $r_B$) increases, the corresponding electroweak gauge boson becomes lighter. Thus the actual composition of the lightest mass eigenstate sensitively depends on the boundary parameters. In the right panel of Fig.~\ref{A1A2contours} we present the level 1 KK Weinberg angle $\sin^{2}\theta_{W}^{(1)}$ as a function of $r_{W}/r_{B}$ for $R^{-1}= 1 \tev$ assuming $r_{B}/L=0.5$.

We may classify the whole parameter space by three distinctive regions: 
\begin{enumerate}
\item For $k_1^B\ll k_1^W$ (which occurs if $r_B\gg r_W$), the (11)-element of $M^2_{n,odd}$ is smaller than the (22)-element, but still much larger than the off-diagonal elements. The lightest eigenstate is almost purely $B^{(1)}$ and we have the ``standard'' MUED dark matter candidate.
\item For $k_1^B\gg k_1^W$ (which occurs if $r_B\ll r_W$), the (22)-element is smaller than the (11)-element, but still much larger than the off-diagonal elements. The lightest eigenstate is almost purely $W_{3}^{(1)}$ and we have what is normally referred to as a KK Z DM candidate, which is almost mass degenerate with the $W^{(1)}_3$.\footnote{The $W^{\pm(1)}$ mass always lies in between the masses of the two neutral eigenstates, such that the LKP is always neutral.} 
\item For  $k_1^B= k_1^W$ (which occurs if $r_B= r_W$), the contribution from the $\left(k^{B/W}_1\right)^2$ on the diagonal entries are identical, and as this part is proportional to the unit matrix, it does not contribute to the mixing angle. 
Then, the KK Weinberg angle is identical to the zero mode (and therefore the SM) Weinberg angle. In this case we have a mixture between the $B^{(1)}$ and the $W_{3}^{(1)}$ resulting in electroweak type KK gauge bosons, the lightest of which we call $A_{1}^{(1)}$ and is the DM candidate. 
\end{enumerate}

A notable feature here is that the Weinberg angle is almost always $\theta^{(1)} \approx 0$ or $\pi/2$ except the region of degenerate $r_W/r_B \approx 1$ where the transition takes place (see Fig. \ref{A1A2contours}.) This feature is easily understood as the off-diagonal entries are relatively small ($\lsim \mathcal{O}(v^2)$) compared to the diagonal entries ($\sim \mathcal{O}(1/R^2)$) so that a small difference in $r_B$ and $r_W$ easily induce an abrupt transition of the LKP from $W_3^{(1)}$-like to $B^{(1)}$-like or vice versa.

\begin{figure}[]
\centerline{
\includegraphics[width=.5\textwidth]{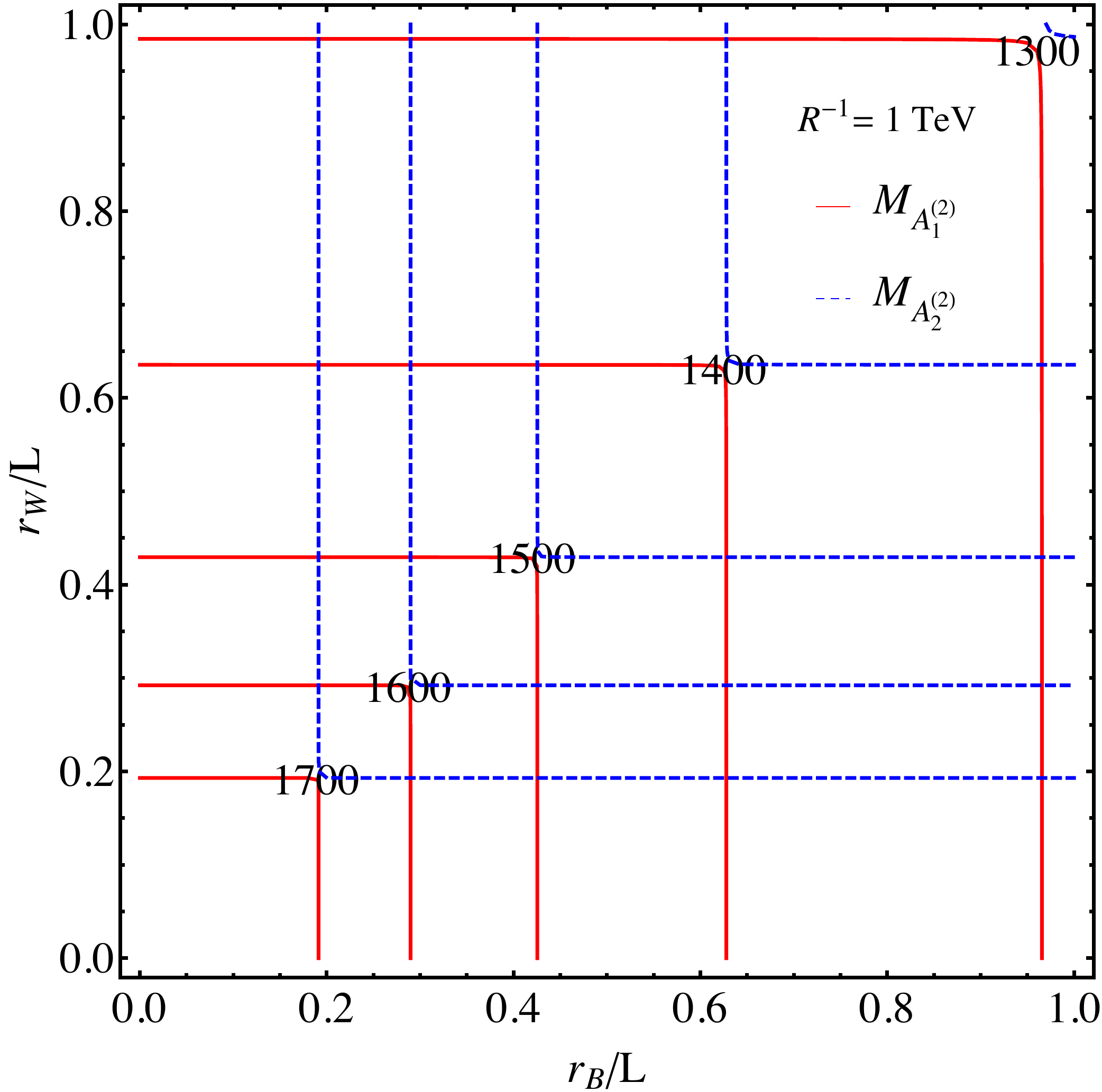} \hspace{0.1cm}
\includegraphics[width=.5\textwidth]{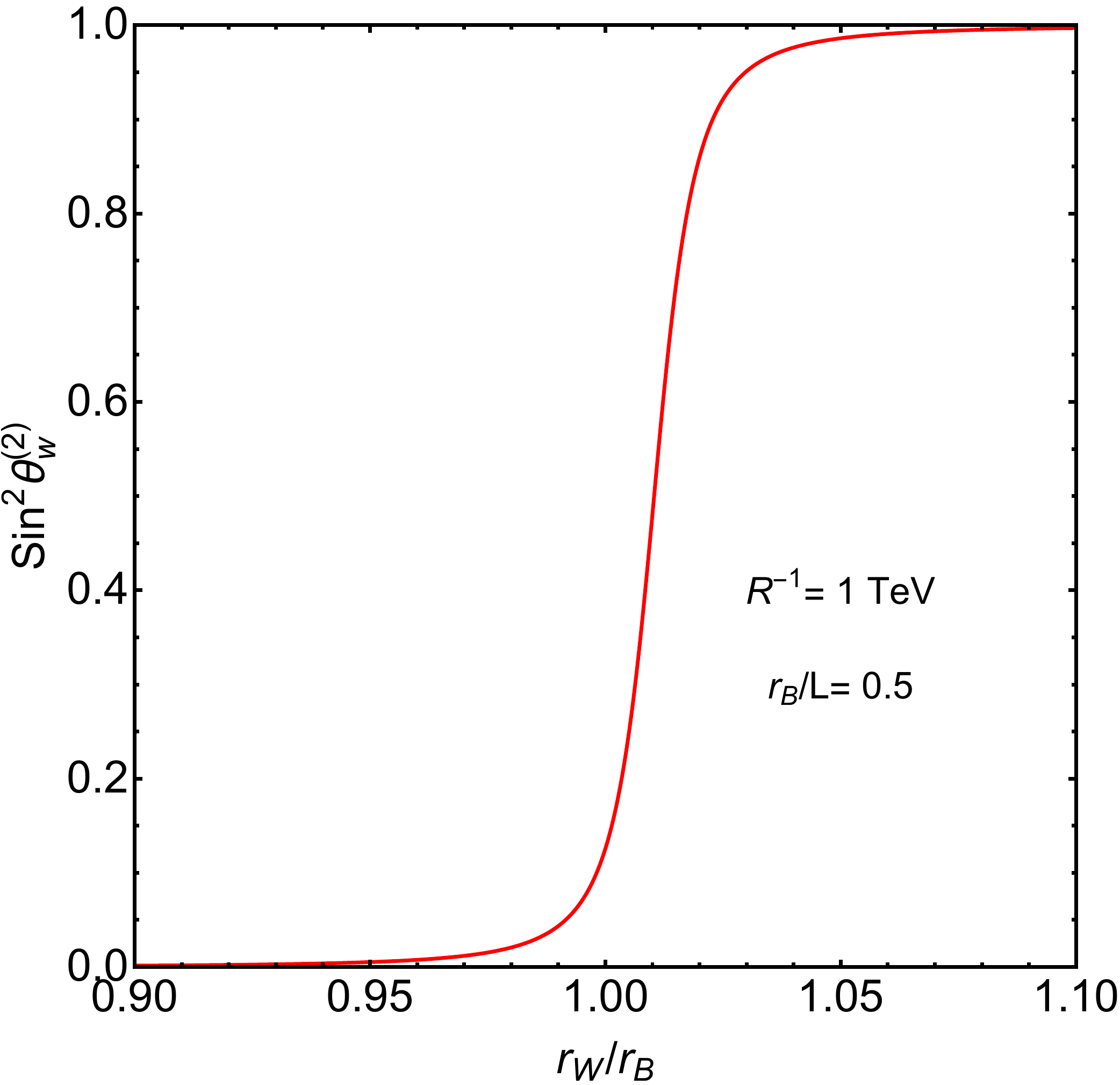} }
\caption{ Left: Contours of constant level 2 gauge boson masses for $R^{-1}=1 $ TeV. The red contours show the mass of the lighter eigenstate $A_{1}^{(2)}$ and the blue contour represents mass of heavier eigenstate $A_{2}^{(2)}$. Right: The level 2 KK Weinberg angle $\sin^2\theta_{W}^{(2)}$ for $R^{-1}$ = 1 TeV and $r_{B}/L=0.5$.  }
\label{GuA}
\end{figure}

The mass spectrum and the properties of the mixing angle of the level 2 KK bosons are analogous to those for the first KK bosons as shown in Fig.~\ref{GuA} (left) where we present the contours for the mass eigenstates $A_{1}^{(2)}$ (the lighter 2nd KK EW boson) and $A_{2}^{(2)}$ (the heavier 2nd KK EW boson), respectively for a fixed compactification scale $R^{-1}= 1$ TeV.  The Weinberg angle of the level 2 bosons, $\sin ^{2}\theta_{W}^{(2)}$,  is depicted in Fig.~\ref{GuA} (right). We can still observe the similar sharp transition near $r_{W}/r_{B}\approx 1$ as is expected from the similar underlying physics in the case for the level 1 EW bosons. We will discuss the detailed phenomenological implications in Sec.~\ref{sec:dilepton}.

\subsection{Coupling between KK bosons and fermions}

The couplings between KK gauge bosons and fermions are determined by a product of the corresponding SM coupling and the wave function overlap integral of the interacting particles ($\AM^{(\ell)}-\psi^{(m)}-\psi^{(n)}$):
\begin{eqnarray}
g_{\AM^{(\ell)}\psi^{(m)}\psi^{(n)}}&=& g_{\AM}\tilde{\FM}_{\ell m n}^{\AM}\\
\tilde{\FM}_{\ell m n}^{\AM}&\equiv&\frac{1}{\mathcal{N}_{0}^{\AM}}\int_{-L}^{L}dy\, f_{\ell}^{\AM}(y)f_{m}^{\psi}(y)f_{n}^{\psi}(y),
\end{eqnarray}
where $g_{\mathcal{A}}$ denotes $g_{1}$ or $g_{2}$, $\mathcal{N}^\AM_{0}$ is normalization factor and $\tilde{\FM}$ are the normalized overlap integrals in Eq.~(\ref{g12def}), which essentially describe the relative strength of the coupling constant with respect to the SM one. All KK number conserving interactions satisfy a `sum-rule' $| \ell \pm m \pm n | = 0 $. However, there are KK number violating interactions which only satisfy the rule from the KK parity conservation: $\ell+m+n \in \mathbb{Z}_{even}$.

Among those couplings, we are first interested in the KK number conserving interactions e.g.,  $\mathcal{A}^{(1)}\bar{\psi}^{(1)}\psi^{(0)}$. This interaction is particularly important in dark matter physics since the dark matter is identified as a level 1 EW gauge boson and it interacts with the SM fermion and its first KK excitation mode with the effective coupling constant 
\begin{eqnarray}
g_{\AM^{(1)}\psi^{(1)}\psi^{(0)}}=g_{\AM}\tilde{\FM}_{110}^{\AM}.
\label{eq:g110}
\end{eqnarray}

\begin{figure}
\centerline{
\includegraphics[width=.5\textwidth]{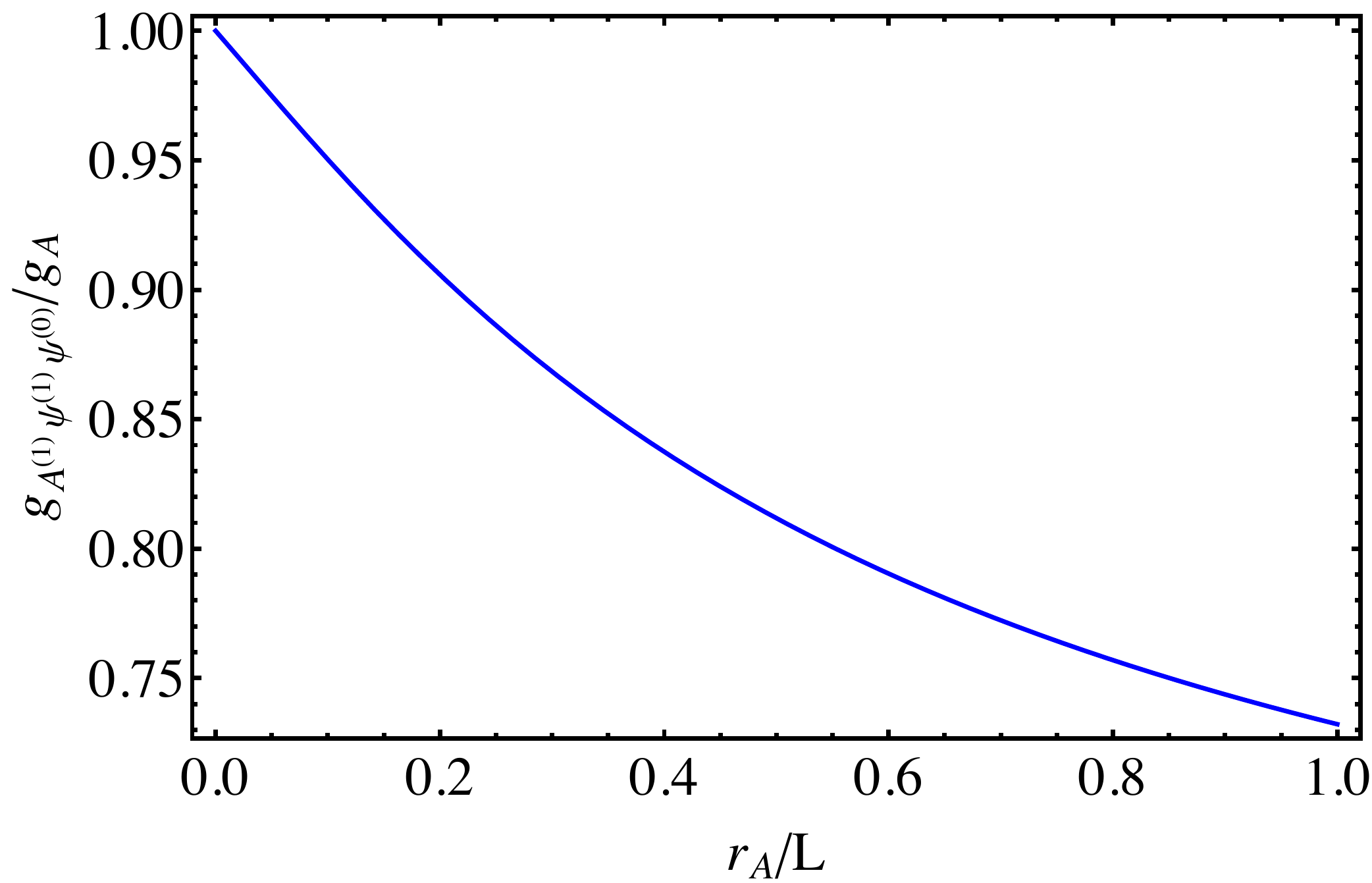}
\includegraphics[width=.5\textwidth]{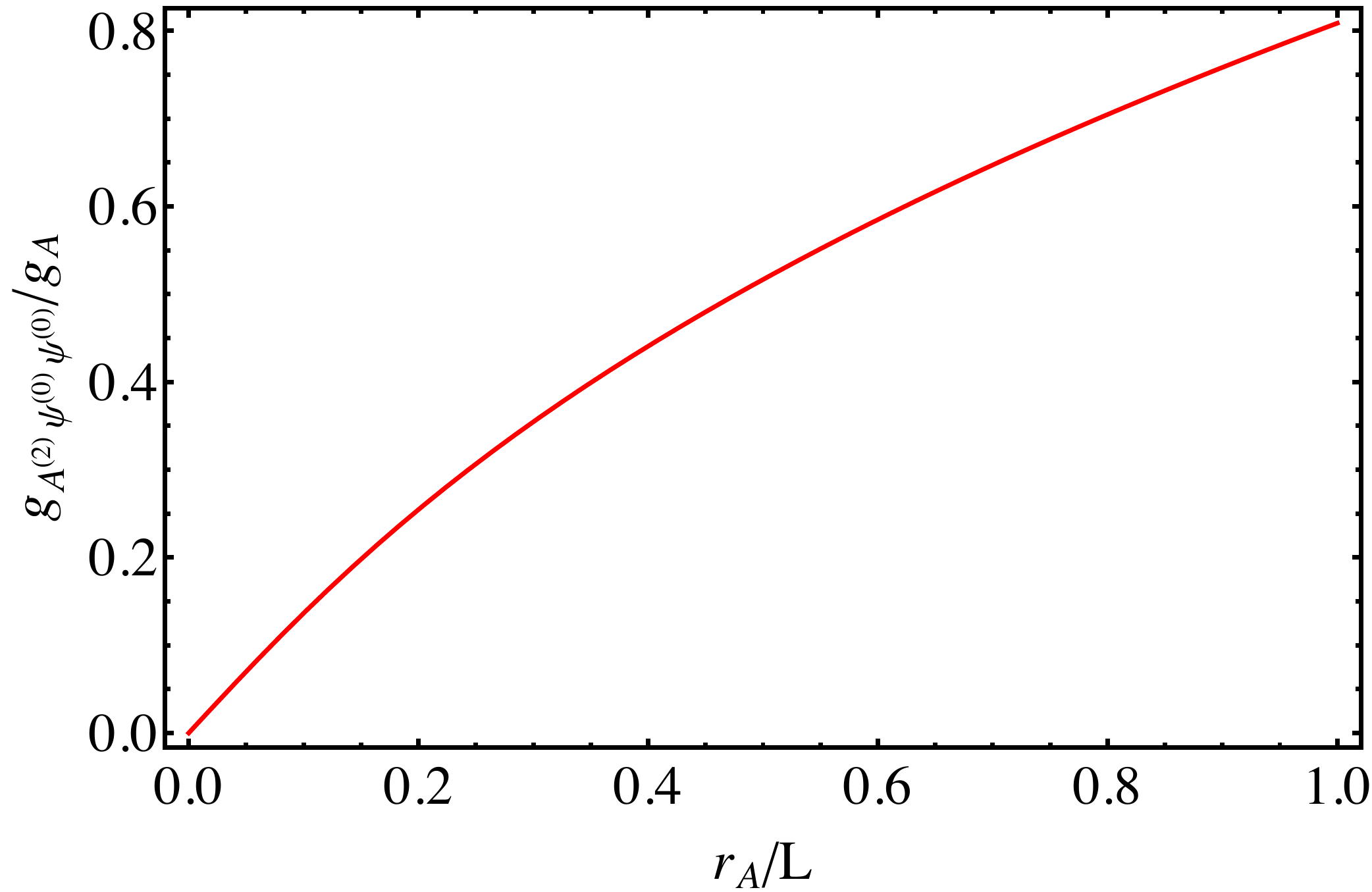}}
\caption{\label{fig:MBvsr} 
Couplings of KK gauge bosons to fermions, normalized with respect to the Standard Model gauge couplings, as a function of the boundary parameter $r_\mathcal{A}$. Left: The KK number conserving coupling of a $U(1)_Y$ or $SU(2)$ first KK mode gauge boson to a first KK mode fermion and a Standard Model fermion ($\tilde{\FM}_{110}^{\AM}$ as defined in Eq.~(\ref{eq:g110})). Right: The KK number violating interaction of a $U(1)_Y$ or $SU(2)$ second KK mode gauge boson to Standard Model fermions ($\tilde{\FM}_{200}^{\AM}$ as defined in Eq.~(\ref{eq:g2n00})).}
\end{figure}

In Fig.~\ref{fig:MBvsr} (left) we plot the effective couplings $g_{B^{(1)}\psi^{(1)}\psi^{(0)}}$ and $g_{W^{(1)}\psi^{(1)}\psi^{(0)}}$ with respect to the SM gauge couplings varying BLKT parameters $r_{B}$ and $r_{W}$ in the parameter range $(0, L)$. The (110) couplings are reduced when the BLKTs get large. 

There are also KK number violating but KK-parity conserving interactions induced by the couplings, e.g. :
\begin{eqnarray}
g_{\AM^{(2n)}\psi^{(0)}\psi^{(0)}}   
=g_{\AM}\tilde{\FM}_{(2n)00}^{\AM},
\label{eq:g2n00}
\end{eqnarray}
which are absent (at tree level) when the boundary terms vanish. 

The non-vanishing couplings could be probed by the precision electroweak precision measurements and collider experiments: the even mode KK bosons mediate four fermion interactions via $t$-channel as well as $s$-channel diagrams. The induced four Fermi operators are subject to the on-going and future precision measurements.  More directly, when BLKTs are sizable, the second level KK gauge bosons are to be produced in high energy collisions with sizable cross sections and they may appear as new heavy $Z^{\prime}$-like resonances at the LHC. This is subject to resonance searches.

Before studying the phenomenological implications of BLKTs we wish to comment on the parameter choice in 
our scenario discussed here, as compared to the minimal UED scenario. In MUED, all boundary terms are assumed to be identical to zero at a cutoff scale $\Lambda$: 
$r_{W/B}(\mu=\Lambda)=0$ and induced at low scale  through renormalization group (RG) running from $\Lambda$~\cite{Cheng:2002iz}.
On the other hand, in the scenario we study in this article, we explicitly consider BLKTs as parameters at low scale, such as the first KK mode resonance scale, i.e., $r_{W/B}=r_{W/B}(\mu=k^{W/B}_1)$ for DM phenomenology,  
such that we can directly compare our analysis with the low energy observables. 
As we are mostly interested in the lightest KK electroweak gauge bosons as a DM candidate in our study, we only consider the effects of $r_{W/B}$ but one may straightforwardly generalize our study by taking  the non-vanishing BLKTs for fermions or the Higgs fields, which we reserve for the future.

\section{Bounds from contact interactions and resonance searches}
\label{sec:bounds}

In this section we consider experimental constraints on the BLKTs using electroweak precision data, namely from four Fermi-operators as well as results from resonance searches at the LHC.

\subsection{4-Fermi interactions}
\label{sec:4-fermi}

Electroweak precision tests (EWPT) provide stringent constraints on low scale KK masses \cite{Appelquist:2002wb, Gogoladze:2006br, Baak:2011ze}.
In the presence of BLKTs, in particular, the KK electroweak (EW) gauge bosons would have tree level couplings with  the SM fermions through KK-number violating but KK-parity conserving couplings so that they contribute to the four Fermi contact operators below the KK scale~\cite{Cho:1997kf,Olive:2016xmw}. It is convenient to parameterize the four Fermi operators following Ref.~\cite{Huang:2012kz}:
\begin{eqnarray}
\LM_{eff}\supset \sum_{f1,f2}\sum_{A,B=L,R}\eta^s_{f_{1}f_{2},AB}\frac{4\pi}{(\Lambda^s_{f1f2,AB})^{2}}\bar{f}_{1,A}\gamma^{\mu}f_{1,A}\bar{f}_{2,B}\gamma_{\mu}f_{2,B},
\end{eqnarray}
where $f_{1,2}$ are fermions (leptons or quarks), $\eta^s_{f1f2,AB}=\pm 1$ and $s=\pm$ are parameters for specific interaction patterns.
The effective cutoff scale is given as
\begin{eqnarray}
\frac{4\pi}{\Lambda^2_{eq,AB}}\eta_{eq,AB}&=&4\pi N_{c}\left[\sum_{n=1}^{\infty}(\tilde{\FM}_{2n00}^{B})^{2}\frac{3}{5}\frac{\alpha_{1}Y_{eA}Y_{qB}}{Q^{2}-M_{B^{(2n)}}^{2}}+\sum_{n=1}^{\infty}(\tilde{\FM}_{2n00}^{W})^{2}\frac{\alpha_{2}T_{eA}^{3}T_{qB}^{3}}{Q^{2}-M_{W_{3}^{(2n)}}^{2}}\right]\nonumber\\
&\approx&-12\pi\left[\sum_{n=1}^{\infty}(\tilde{\FM}_{2n00}^{B})^{2}\frac{3}{5}\frac{\alpha_{1}Y_{eA}Y_{qB}}{M_{B^{(2n)}}^{2}}+\sum_{n=1}^{\infty}(\tilde{\FM}_{2n00}^{W})^{2}\frac{\alpha_{2}T_{eA}^{3}T_{qB}^{3}}{M_{W_{3}^{(2n)}}^{2}}\right].
\label{4fermieq} 
\end{eqnarray}

\begin{table}[b]
\caption{Four Fermi contact interaction bounds in TeV from PDG (2016)  \cite{Olive:2016xmw}}
\begin{center}
\begin{tabular}{c|c|c|c|c|c|c|c}
TeV&$eeee$& $ee\mu\mu$&  $ee\tau\tau$& $\ell\ell\ell\ell$&$qqqq$&$eeuu$&$eedd$ \\    
\hline
\hline
$\Lambda_{LL}^+$ &$>8.3$&$>8.5$&$>7.9$&$>9.1$&$>9.0$&$>23.3$&$>11.1$ \\
$\Lambda_{LL}^-$  &$>10.3$&$>9.5$&$>7.2$&$>10.3$&$>12.0$ &$>12.5$&$>26.4$
\end{tabular}
\end{center}
\label{t:4fermibds}
\end{table}

The effective couplings are weighted by the factor $\tilde{\FM}_{2n00}^{\AM}$ which is the integrated wave function overlaps from $\AM^{(2n)}\bar{\psi}^{(0)}\psi^{(0)}$ couplings in Eq.~(\ref{eq:g2n00}) with a color factor $N_c=3$. The quantum numbers $Y$'s and $T$'s are the hypercharges and isospins of the interacting fermions (electron and quarks) and 
we take the one-loop improved values
\begin{eqnarray}
\alpha_{1}(\mu)&=&\frac{5}{3}\frac{g_{Y}^{2}(\mu)}{4\pi}=\frac{\alpha_{1}(m_{z})}{1-\frac{b_{1}}{4\pi}\alpha_{1}(m_{Z})\log\frac{\mu^{2}}{m_{Z}^{2}}},\nonumber\\
\alpha_{2}(\mu)&=&\frac{g_{ew}^{2}(\mu)}{4\pi}=\frac{\alpha_{2}(m_{Z})}{1-\frac{b_{2}}{4\pi}\alpha_{2}(m_{Z})\log\frac{\mu^{2}}{m_{Z}^{2}}},
\end{eqnarray}
where $\alpha_{1}(m_{Z})\approx 0.017, \alpha_2{(m_{Z})}\approx 0.034$, and $(b_{1},b_{2})=(41/10, -19/6)$ below the compactification scale.

\begin{figure}[t]
	\centering
		\includegraphics[width=.48\textwidth]{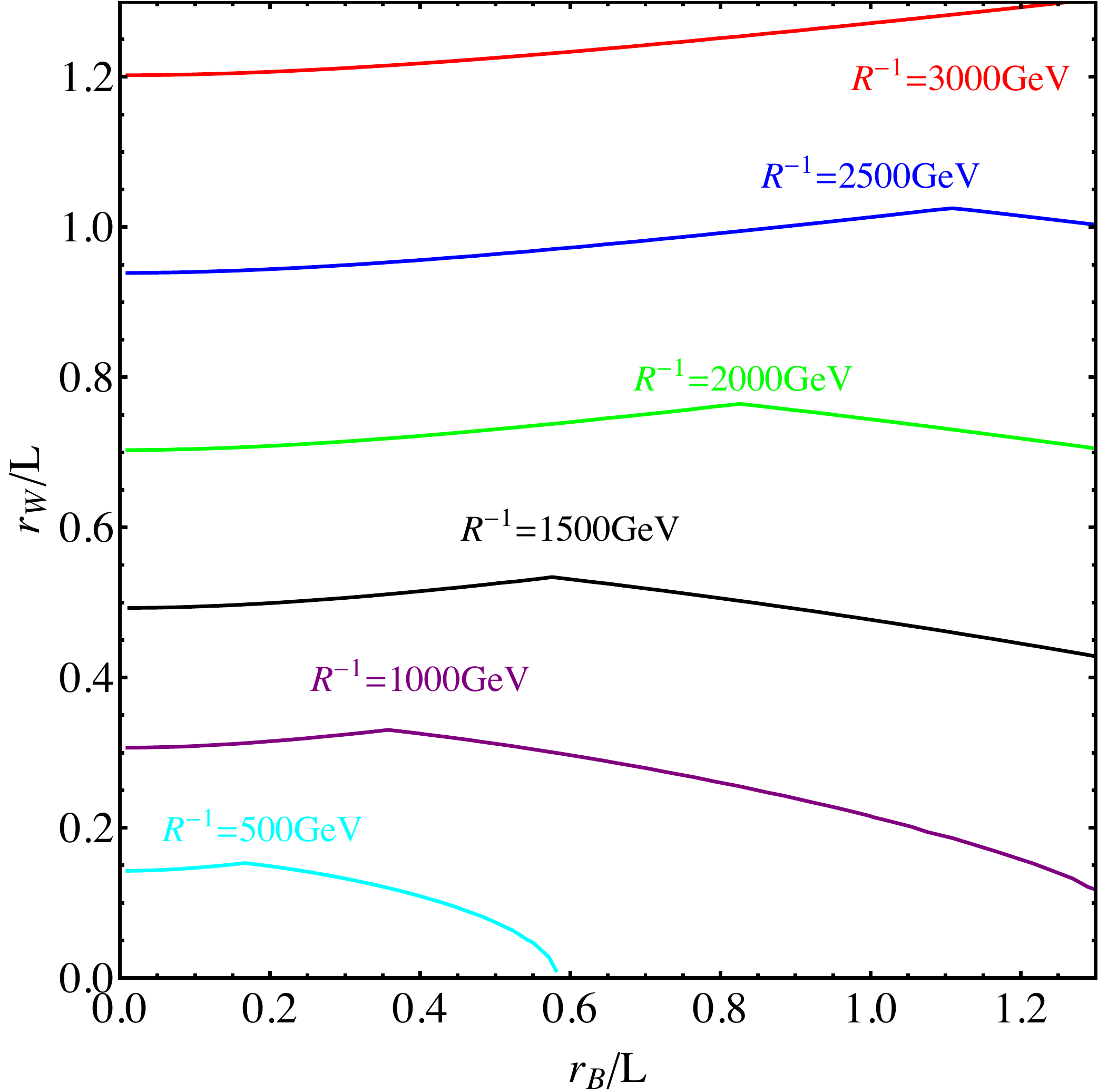}	\hspace*{0.2cm}
		\includegraphics[width=.48\textwidth]{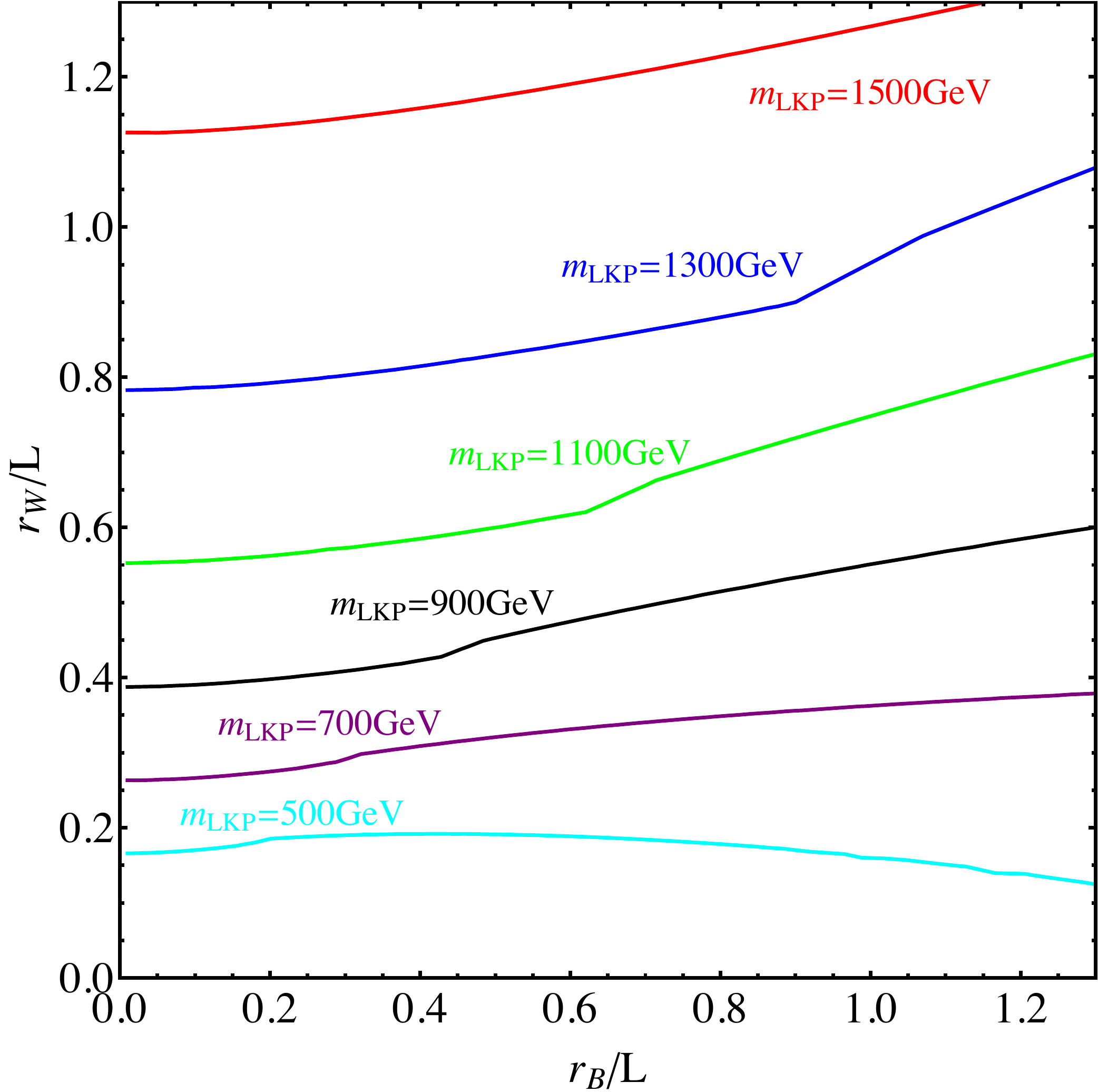}
	\caption{\label{4fermi} 
		Bounds on $R^{-1}$ from four-Fermi contact interactions in $r_{B}/L$ and $r_{W}/L$ space. Left: Contours of minimally allowed values of $R^{-1}$. Right: Contours of minimally allowed values of $m_{LKP}$. }
		\end{figure}

Equipped with the effective parameterization of four Fermi operators, we are now ready to compare with the experimental results. 
We take the updated results in PDG 2016 \cite{Olive:2016xmw} as the reference values of experimental bounds.
We summarize the relevant results in Table~\ref{t:4fermibds}. The most stringent bound arises from the $e_{L}e_{L}q_{L}q_{L}$ interaction.  Fig.~\ref{4fermi} shows the bounds on $R^{-1}$ in the $(r_W/L, r_B/L)$ plane. The most important ones are the results from $eeuu$ and $eedd$. We draw the contours for various values of $R^{-1}\in (500, 3000)$ GeV (left) and the LKP mass, $m_{LKP}\in(500, 1500)$ GeV (right). The region above the line with a given $R^{-1}$ (or $m_{LKP}$) is ruled out (thus the region below the line is allowed) because the effective couplings are too large. 
As expected, a larger parameter space is allowed for a large $R^{-1}$ (and $m_{LKP}$) because of large suppression factors ($\sim 1/m_{LKP}^2$) in the effective operators.  We notice that the bounds are more sensitive to the boundary parameter $r_{W}$ rather than $r_B$ mainly due to the large weak coupling compared to the hypercharge coupling. For example, above $m_{LKP} \approx 700 ~(1100)$ GeV, essentially no stringent bound is found on $r_B/L$ but only a restricted region $r_W/L \lsim 0.3-0.4~(0.6-0.8)$  is allowed.

\subsection{Dilepton resonance searches}
\label{sec:dilepton}

The resonance searches at colliders are an effective way of probing BLKTs since  the KK number violating interactions, e.g., $g_{\AM^{(2)}\psi^{(0)}\psi^{(0)}}$ in Eq. (\ref{eq:g2n00}), allow the single production of the second KK gauge bosons at particle collisions.  When the 2nd EW gauge bosons are produced in high energy collisions, they decay to the SM particles which can be observed as a resonance. The production cross sections and the decay widths are all determined by the KK number violating couplings in Eq.~(\ref{eq:g2n00}). Here we focus in particular on dilepton resonances  because the Standard Model background is very low and the expected production cross sections are sizable. The dijet final state has a larger cross section in NMUED, but it is accompanied by a huge QCD background.

To determine the relevant couplings for mass eigenstates $A_{1,2}^{(2)}$ to the standard model fermions, we need to diagonalize the mass matrix in Eq.~(\ref{KKevenMass}). 
The mass matrix at the second KK mode level  reads 
\beq
M^2_{n,2}=\left(
\begin{array}{cc}
(k^B_2)^2+\frac{g_{1}^{2}v^{2}}{4} \tilde{\mathcal{F}}^{BB}_{22} & \frac{g_{1}g_{2}v^{2}}{4}\tilde{\mathcal{F}}^{BW}_{22}\\
 \frac{g_{1}g_{2}v^{2}}{4}\tilde{\mathcal{F}}^{BW}_{22} & (k^W_2)^2+\frac{g_{2}^{2}v^{2}}{4} \tilde{\mathcal{F}}^{WW}_{22}
\end{array}
\right),
\eeq 
where $k^{B,W}_2$ follow from the mass quantization condition in Eq.~(\ref{gaugeKKmass}). Wave function overlaps for $\AM(\AM') =B$ or $W$ defined in Eq.~(\ref{Fnorm}) are
\bea
\tilde{\mathcal{F}}^{\AM\AM'}_{22}&=&   \,\,\sqrt{\frac{1+r_\AM/L}{1+\frac{r_{\AM}}{L}\cos^2(k^\AM_2 L)}}\,\,\sqrt{\frac{1+r_{\AM'}/L}{1+\frac{r_{\AM'}}{L}\cos^2(k^{\AM'}_2 L)}}  \nonumber \\
&&~~~~~~~~~\times \left[\frac{\sin((k^\AM_2+k^{\AM'}_2) L)}{(k^\AM_2+k^{\AM'}_2) L}+\frac{\sin((k^\AM_2-k^{\AM'}_2) L)}{(k^\AM_2-k^{\AM'}_2) L}\right].
\eea
For $\AM=\AM'$, the second term in the square parenthesis becomes 1:
\bea
 \tilde{\mathcal{F}}^{WW}_{22}&=&\,\,\frac{1+r_W/L}{\frac{r_W}{L}\cos^2(k^W_2 L)}\,\,\left[\frac{\sin(2k^W_2 L)}{2k^W_2 L}+1\right],\\
 \tilde{\mathcal{F}}^{BB}_{22}&=&\,\,\frac{1+r_B/L}{1+\frac{r_B}{L}\cos^2(k^B_2 L)}\,\,\left[1+\frac{\sin(2k^B_2 L)}{2k^B_2 L}\right].
\eea

The mass matrix is diagonalized by a rotation by an angle $\theta_W^{(2)}$, the weak rotation angle for the 2nd level KK gauge bosons:
\begin{eqnarray}
\begin{pmatrix}
A_{1}^{(2)} \\
A_{2}^{(2)}
\end{pmatrix}=U_{n}^{(2)}\begin{pmatrix}
B^{(2)}\\
W_{3}^{(2)}
\end{pmatrix}=\begin{pmatrix}
\cos(\theta_{W}^{(2)}) & \sin(\theta_{W}^{(2)})\\
-\sin(\theta_{W}^{(2)}) & \cos(\theta_{W}^{(2)})
\end{pmatrix}\begin{pmatrix}
B^{(2)}\\
W_{3}^{(2)}
\end{pmatrix}.
\end{eqnarray}

Interaction of the mass eigenstates and zero mode fermions are obtained from 
interaction of gauge eigenstates ($W^{(2)}_{\pm,3}$ and $B^{(2)}$), which follows from the covariant derivative,
\bea
\mathcal{L}\supset \int dy \bar{\Psi} i\slashed{D} \Psi& \supset & - g_{W^{(2)}\psi^{(0)}\psi^{(0)}}\bar{\psi}^{(0)} W^{(2)}_3 T^3_{\psi_L} P_L \psi^{(0)}\nonumber\\
&&-g_{B^{(2)}\psi^{(0)}\psi^{(0)}}\bar{\psi}^{(0)} B^{(2)} (Y_{\psi_L} P_L +Y_{\psi_R} P_R)  \psi^{(0)}\nonumber\\
&=& - g_{W^{(2)}\psi^{(0)}\psi^{(0)}}\bar{\psi}^{(0)} \left(\sin(\theta^{(2)}_W) A^{(2)}_1 +  \cos(\theta^{(2)}_W) A^{(2)}_2 \right) T^3_{\psi_L} P_L \psi^{(0)}\\
&&-g_{B^{(2)}\psi^{(0)}\psi^{(0)}}\bar{\psi}^{(0)}\left(\cos(\theta^{(2)}_W) A^{(2)}_1 -  \sin(\theta^{(2)}_W) A^{(2)}_2 \right)  (Y_{\psi_L} P_L +Y_{\psi_R} P_R)  \psi^{(0)},\nonumber
\eea
where 
\bea
g_{W^{(2)}\psi^{(0)}\psi^{(0)}}&=&g_2\int_{-L}^{L} \frac{dy}{2L} \frac{f^W_2(y)}{\mathcal{N}^W_0}=g_2 \sqrt{\frac{2(1+r_W/L)}{1+\frac{r_W}{L}\cos^2(k^W_2 L)}}\frac{\sin(k^W_2 L)}{k^W_2 L},\label{overlap2n00_1}\\
g_{B^{(2)}\psi^{(0)}\psi^{(0)}}&=&g_1\int_{-L}^{L} \frac{dy}{2L} \frac{f^B_2(y)}{\mathcal{N}^B_0}=g_1 \sqrt{\frac{2(1+r_B/L)}{1+\frac{r_B}{L}\cos^2(k^B_2 L)}}\frac{\sin(k^B_2 L)}{k^B_2 L}.
\label{overlap2n00_2}
\eea

\begin{figure}[t]
\centering
\includegraphics[width=0.6\textwidth]{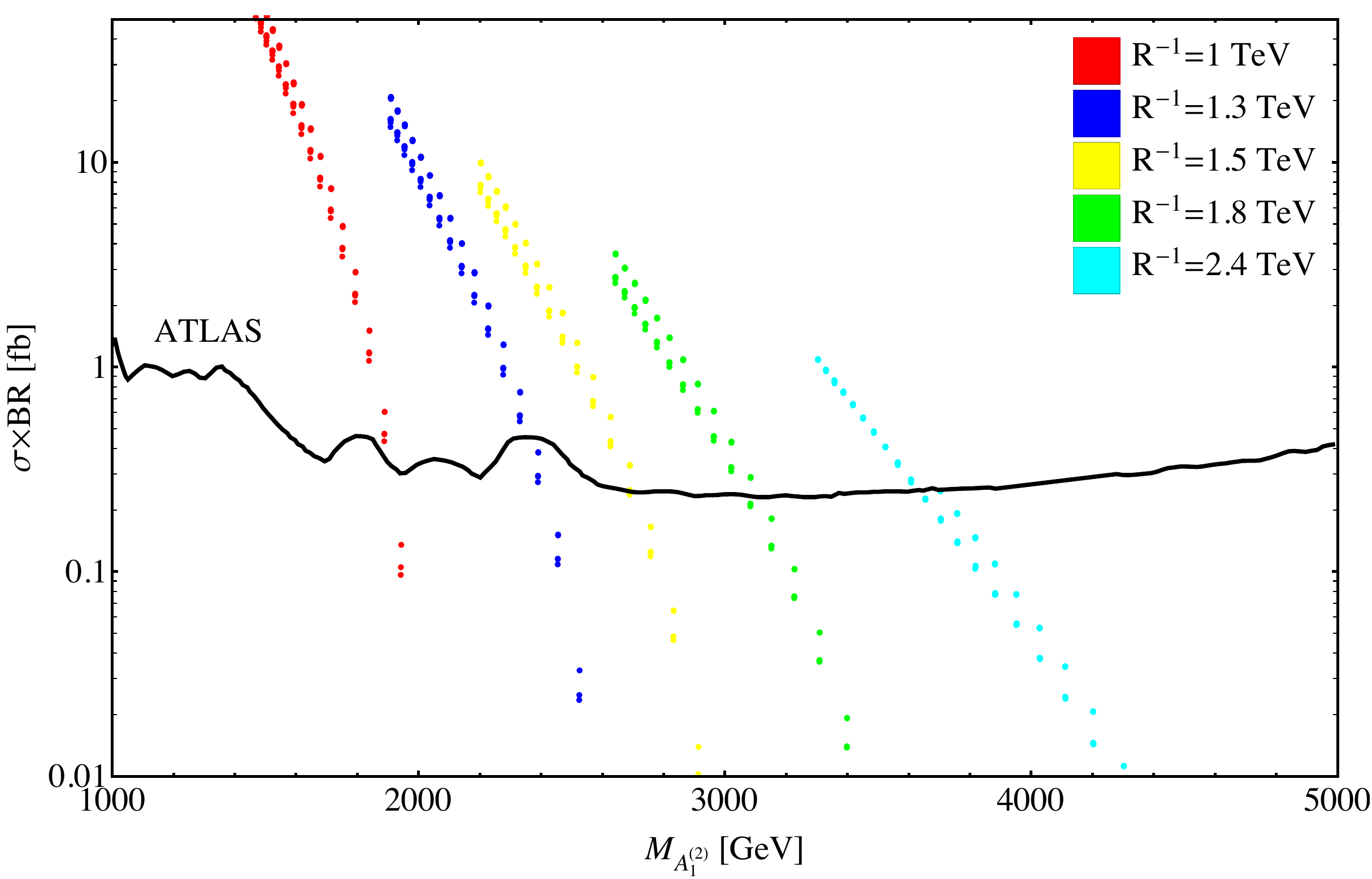}
\includegraphics[width=0.377\textwidth]{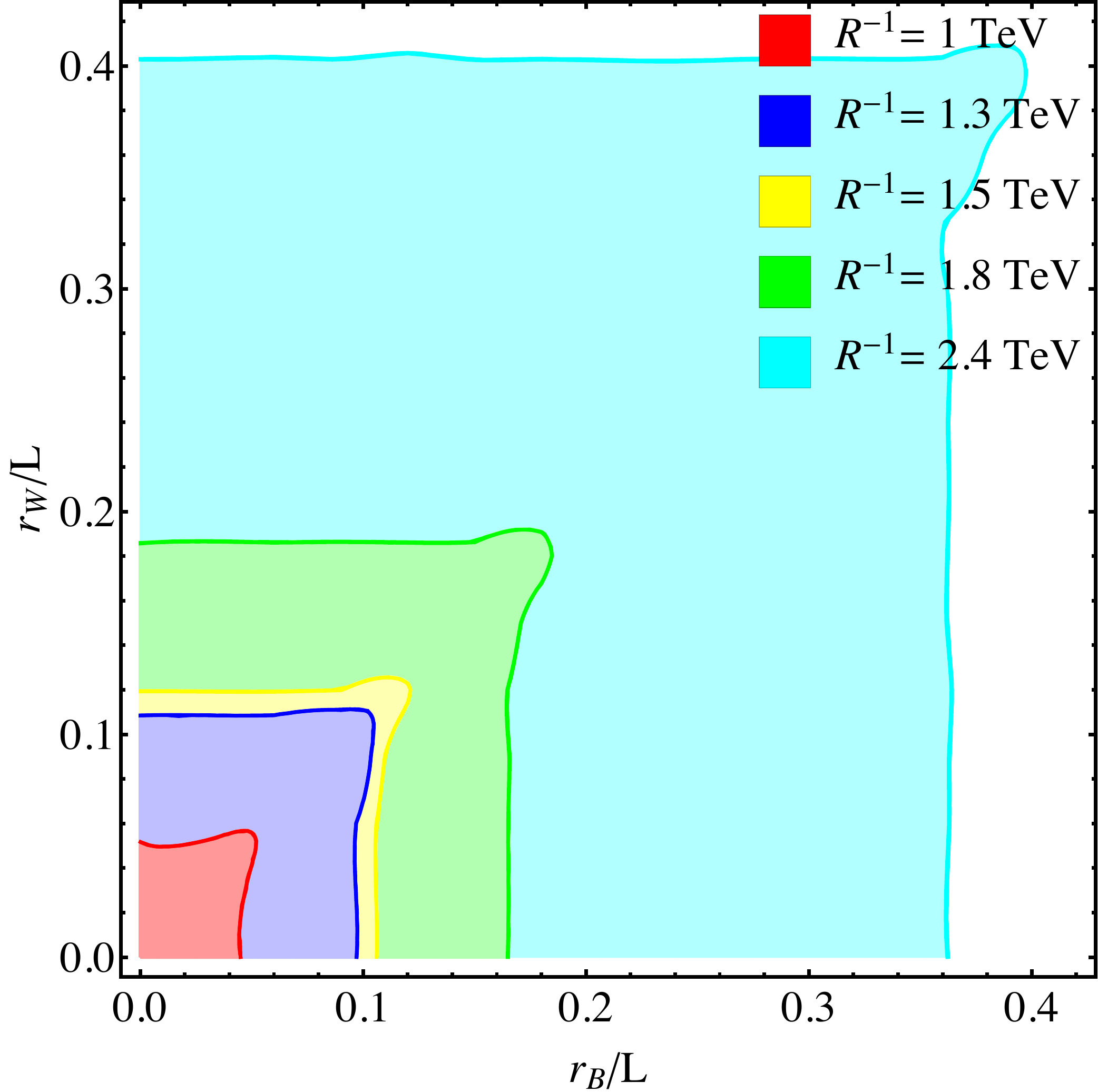}
\caption{
Constraints from  the 13 TeV ATLAS dilepton resonance search with 13.3 fb$^{-1}$ luminosity \cite{ATLAS:2016cyf}, and bounds on the NMUED parameter space. Left: Constraints on the cross section times branching ratio to two leptons by ATLAS as a function of the resonance mass (black, solid).  The model predictions for $A^{(2)}_1$ (lighter) resonance signals with  $R^{-1}=1, 1.3, 1.5, 1.8, 2.4$ TeV in the parameter window $(r_B/L \in (0,1), r_W/L\in(0,1))$ are given by scatter plots. Low values of $r_B/L$ and $r_W/L$ correspond to low cross sections. 
Right: Bounds on the NMUED parameter space from dilepton searches in the $r_W/L$ vs. $r_B/L$  plane. The red shaded region shows the allowed parameter space assuming $R^{-1} = 1 ~{\rm TeV}$. The blue region is for  $R^{-1} = 1.3 ~{\rm TeV}$, the yellow region for $R^{-1} = 1.5 ~{\rm TeV}$, the green region for $R^{-1} = 1.8 ~{\rm TeV}$ and the larger cyan region is for $R^{-1} = 2.4 ~{\rm TeV}$.
}
\label{dileptonres}
\end{figure}

The ATLAS and CMS collaborations have searched for heavy narrow dilepton resonances at 13 TeV with 13.3 fb$^{-1}$ (ATLAS)~\cite{ATLAS:2016cyf} and 13.0 fb$^{-1}$(CMS)~\cite{CMS:2016abv} data, respectively.  
The experimental bounds are set in the combination of the production cross section of the heavy resonance particle and the branching fraction to dileptons, $\sigma\times BR(\ell \bar \ell)$.  The bounds are similar in both experiments. Here we use the ATLAS results, which are based on a slightly larger set of data. 
Since the relevant production cross sections and the branching fractions are given by three parameters $r_{B}/L, r_{W}/L$ and $R^{-1}$, we find the allowed parameter space in $(r_{B}/L, r_{W}/L)$ for various values of $R^{-1}$. In the left panel of Fig.~\ref{dileptonres} we show the ATLAS upper limit on $\sigma \times BR$ in the mass range $(1000, 5000)$ GeV and the expectations for the lighter level-2 KK gauge boson $A^{(2)}_1$ decaying to leptons. With a large compactification scale $R^{-1}$, a heavy dilepton resonance is expected so that a large parameter space in $(r_{B}/L, r_{W}/L)$ is allowed as shown in the right panel.  For $R^{-1}=2.4$ TeV, roughly $r_\AM/L\lsim 0.4$ is allowed for $\AM=W, B$ but a smaller $R^{-1}=1.5$ TeV for instance is compatible only with a smaller range $r_\AM \lsim 0.2$ or so.
When comparing the results presented in Fig.~\ref{dileptonres} and Fig.~\ref{4fermi}, the LHC bound from the dilepton search is by far more stringent than the results from four-Fermi contact interactions. 
For the calculation of signal cross sections at leading order ($pp \to \AM \to \ell \bar\ell$), we have used {\tt CalcHEP} \cite{Belyaev:2012qa} and {\tt MG5\_aMC@NLO} \cite{Alwall:2014hca} with masses and couplings defined above. 

While a dedicated study with double narrow resonance may provide more stringent bounds, we include the lighter level-2 KK gauge boson, $A^{(2)}_1$, in our analysis, 
since current ATLAS/CMS analysis assumes a single resonance in the dilepton channel. 
We assume that level-2 KK gauge bosons dominantly decay into SM fermion final states, and the decay width is computed automatically while scanning over $(r_{B}/L, r_{W}/L)$ for a given $R^{-1}$. Similar or slightly weaker bounds are obtained with the heavier KK gauge boson, $A^{(2)}_2$.

\section{Phenomenology of electroweak KK DM}
\label{sec:DM}

Conventionally the KK photon has been regarded as a dark matter candidate in the literature. Here we focus on an LKP formed from a mixture of the first KK excitation of the hypercharge gauge boson
and the neutral component of the $\rm SU(2)_W$ gauge boson. We have illustrated that when BLKTs are involved, mixing can show interesting features. In this section, we examine the phenomenology
of the mixed LKP, dubbed as electroweak KK dark matter, while considering existing bounds on the BLKTs as discussed in the previous sections.

\subsection{Relic abundance}

In section \ref{sec:setup0}, we investigated the mass spectra and couplings of the KK electroweak gauge bosons in the presence of BLKTs. The masses, couplings and mixing angles sensitively depend on the BLKT parameters $r_{B}$ and $r_{W}$. 
Therefore the annihilation cross-sections and the relic density ($\sim 1/\langle \sigma v \rangle$) are affected as well. 
The relevant interactions are
\begin{eqnarray}
\AM^{(1)}\psi^{(1)}\psi^{(0)} &&\text{KK boson-KK fermion-SM fermion} \, ,\\
\AM^{(1)}\AM^{(1)}H^{(0)}H^{(0)} &&\text{KK boson-KK boson-Higgs-Higgs} \, ,\\
\AM^{(1)}\AM^{(1)}\AM^{(0)} && \text{KK boson-KK-boson-SM boson} \, ,
\end{eqnarray}
where $\AM$ collectively stands for the mass eigenstate of the KK electroweak gauge boson $A_{1}$ or $A_{2}$. As it is clearly seen in the left panel of Fig.~\ref{fig:MBvsr}, the couplings monotonically decrease as a function of BLKT parameters.

We compute the relevant couplings and identify DM candidate from the mass eigenstates, and then rescale MUED couplings for annihilation cross sections in Ref. \cite{Kong:2005hn}. 
KK fermions and KK Higgs masses are set to $\sqrt{ \left ( \frac{1}{R} \right )^2 + m_{SM}^2}$, since we include no  boundary terms for them. For $r_W \sim r_B$, 
$A^{(1)}_{1}$ and $A^{(1)}_{2}$ are degenerate and therefore coannihilation processes are important. Since the mass of $W^{(1)\pm }$ is always between those of the two neutral gauge bosons, 
we include coannihilation processes with $A^{(1)}_{2}$, and $W^{(1)\pm }$ in addition to self-annihilation of $A^{(1)}_{1}$. 

\begin{figure}[]
\begin{center}
  \centering
    \includegraphics[width=0.6\textwidth]{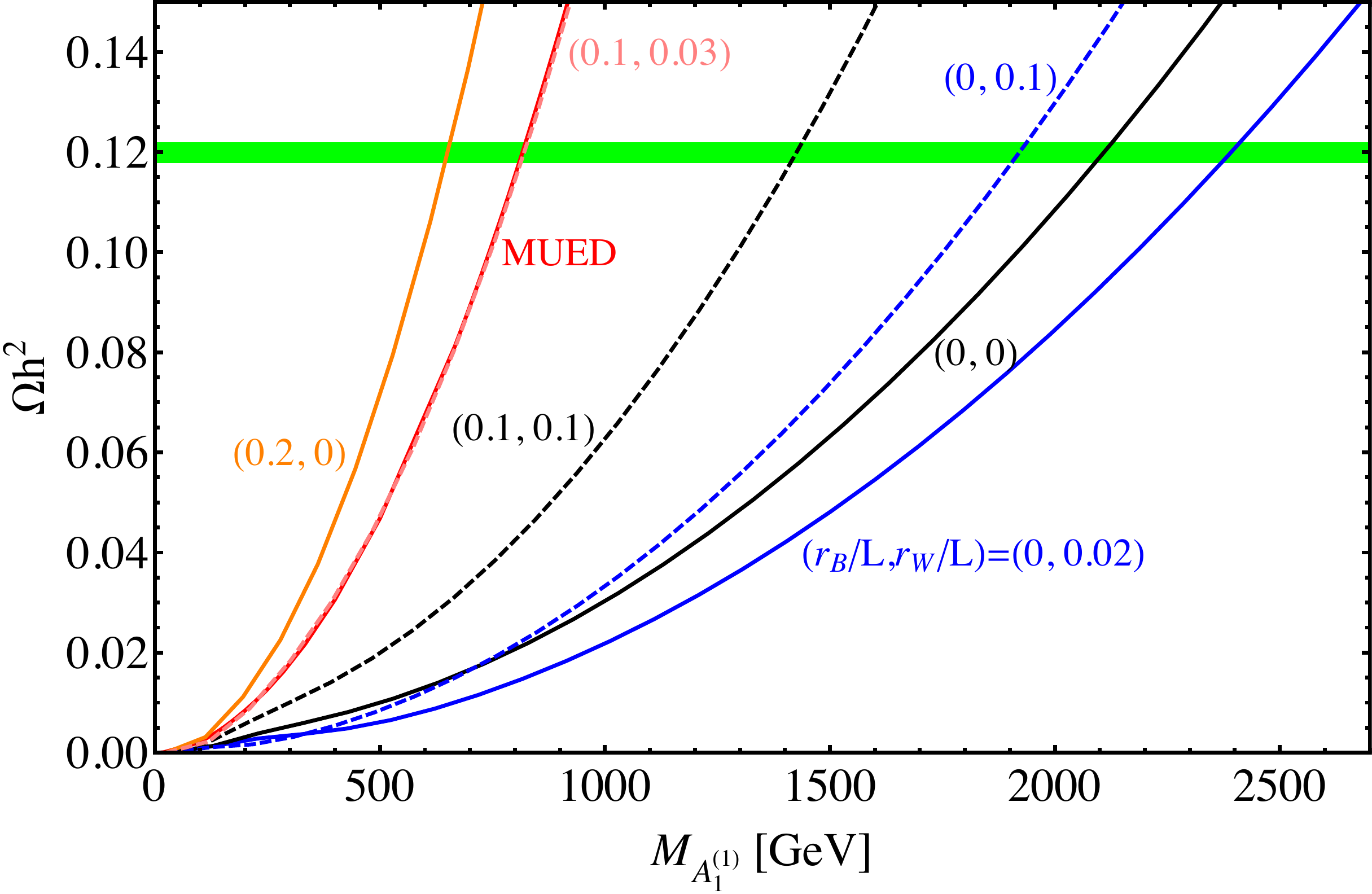}
  \caption{\label{relic_dm_mass}
    Relic density $\Omega h^{2}$ as a function of DM mass $M_{A_{1}^{(1)}}$ for given values of $r_{B}/L$ and $r_{W}/L$. Different colors and line-styles indicate different ($r_{B}$, $r_{W}$) values. The green band indicate $\Omega h^{2}$ from the result of Planck 2015~\cite{Ade:2013zuv}.} 
    \end{center}
\end{figure}

Our results are shown in Fig.~\ref{relic_dm_mass} as a function of the DM mass for various values of $(r_{B}/L, r_{W}/L)$. The red curve presents the relic density of the conventional DM candidate in MUED, which is the hypercharge gauge boson \cite{Servant:2002aq, Arrenberg:2008wy}. In this case, the dominant annihilation final states are SM fermions with a small contribution from the Higgs-Higgs final state. 
With 1-loop corrected mass spectrum, the Weinberg angle at level-1 is very small and therefore there is no gauge boson final state \cite{Cheng:2002iz}.
We find the MUED results are reproduced with $(r_{B}/L, r_{W}/L) = (0.1,0.03)$ which is shown as the pink dashed line.\footnote{The MUED line is understood as follows.  In MUED, it is assumed that boundary parameters are all set to be zeroes as $\left.(r_B/L, r_W/L)\right |_{\rm cut-off}=(0,0)$ at the cut-off scale.  After the renormalization group evolution, the non-zero boundary parameters are radiatively generated at the electroweak (EW) scale. Note again that we set the boundary parameter $(r_{B}/L,r_{W}/L)$ at the electroweak scale in this article unless it is notified differently.}
 The blue solid line shows a parameter choice which yields the maximum allowed value of the DM mass around 2.4 TeV. We indicate the observed relic abundance $\Omega h^{2}$ from the Planck collaboration~\cite{Ade:2013zuv} in the green band.   

Since the KK mixing angle changes rapidly in the vicinity of the line along $|r_{B}-r_{W}|= 0$ as can be seen in Fig.~\ref{A1A2contours}, the DM phenomenology is strongly altered when crossing this parameter region. For $r_{W} > r_{B}$, the main component of the LKP $A_{1}^{(1)}$ becomes  $W_{3}^{(1)}$.  Since the ${\rm SU}(2)$ coupling is stronger than the ${\rm U}(1)_Y$, the annihilation cross section in this regime becomes greater than the value for the $B^{(1)}$-like LKP, which implies that the observed relic density is reproduced at a much larger LKP mass. That is the reason why we get smaller LKP mass for $(r_{B}/L, r_{W}/L)=(0.2, 0)$ shown in the orange line in Fig.~\ref{relic_dm_mass}. However, if $r_{W}/L$ is further increased, the size of effective gauge coupling becomes smaller and the LKP mass has to be reduced to compensate the effect. We find that the allowed upper limit on the LKP mass gets larger up to a critical point $r_{W}/L=0.02$ and then drops down for a larger $r_{W}/L$. This feature is shown as blue solid line and blue dashed line in Fig.~\ref{relic_dm_mass}. The maximally allowed mass of the electroweak gauge boson LKP is about 2.4 TeV, which is significantly higher than the ``naive'' MUED value of $0.9$ TeV for the KK photon LKP in MUED \cite{Arrenberg:2008wy}. 

For $(r_{B}/L, r_{W}/L) = (0, 0.02)$, the masses of neutral KK bosons $A_{1}^{(1)}$, $A_{2}^{(1)}$ are 2.38 and 2.43 TeV, respectively. The lightest charged KK boson $W_{1}^{(1)}$ has a mass only slightly higher than the LKP and its contribution to the coannihilation is important, and is fully taken into account in our analysis. In passing, we would mention the potential enhancement of the annihilation cross section through the resonance of the 2nd KK excitation modes. It has been shown in MUED that  2nd KK resonance effects can greatly enhance the annihilation cross section so that the LKP mass could be as high as $\sim 1.3$ TeV~\cite{Belanger:2010yx}. However, the resonance effect becomes important only when the second KK mode masses ($m_{\rm 2nd}$) are very close to twice the mass of the first KK mode ($m_{\rm 1st}$).  This relation between the first and second KK mode masses is satisfied in UED in the absence of boundary terms, but, as can be seen when comparing Figs.~\ref{A1A2contours}~(left) and~\ref{GuA}~(left), in the presence of BLKTs, the second KK resonance is becoming heavier than twice the first KK resonance. 
%
\begin{figure}[t]
  \centering
    \includegraphics[width=0.6\textwidth]{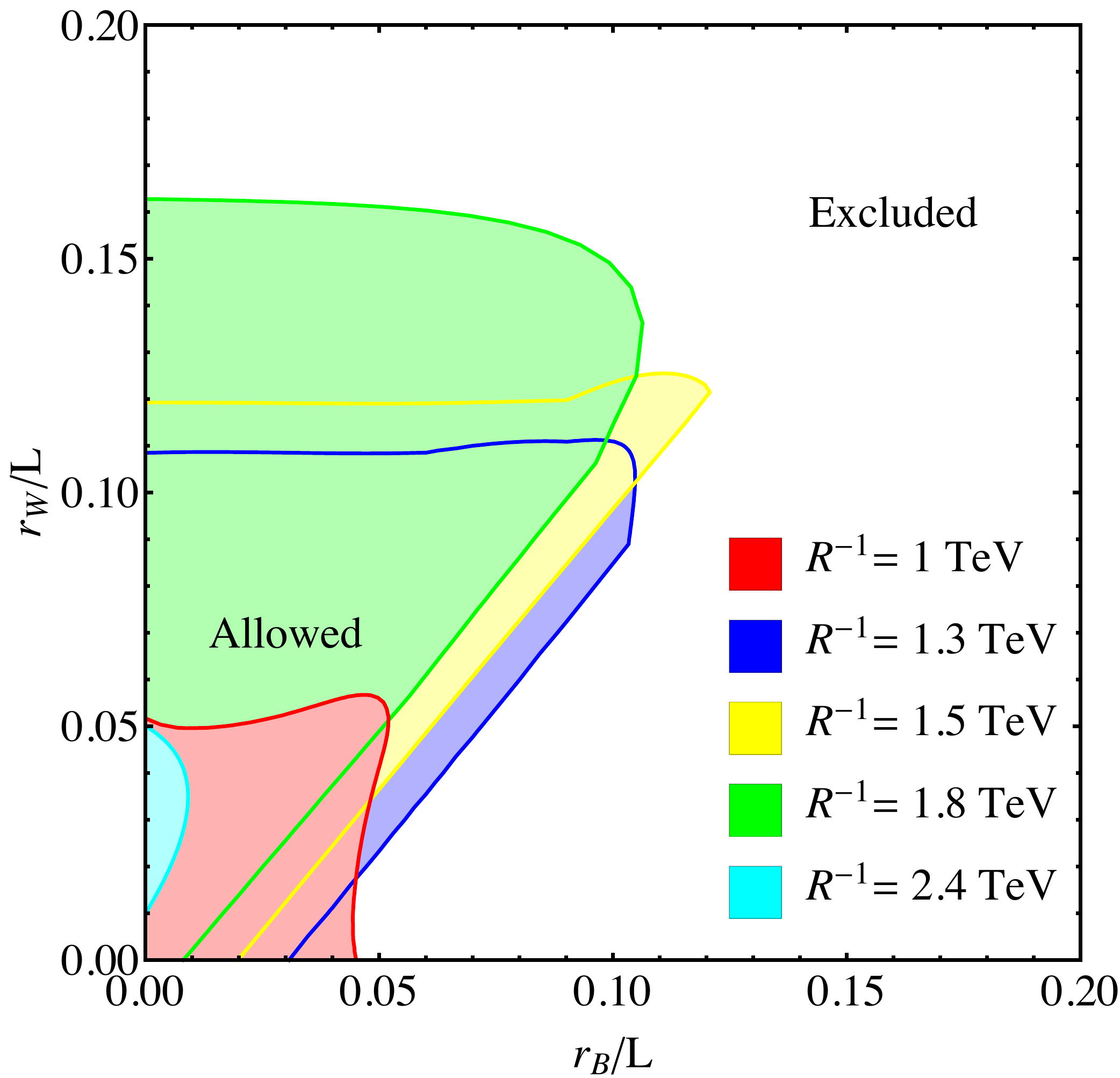}
     \caption{\label{fig:relic_contour}
     {Combined bounds from dilepton searches for $A^{(2)}_{1}$ and $A^{(2)}_{2}$ and from over-closure of the Universe by the relic density of the LKP in the $r_W/L$ vs. $r_B/L$ parameter plane. The shaded regions represent the allowed parameter space for various values of $R^{-1}$.}}
\end{figure}

In Fig.~\ref{fig:relic_contour} we show the 
parameter space for various values of  $R^{-1}\in (1.0, 2.4)$ TeV, which is allowed by dilepton resonance searches (for $A^{(2)}_{1,2}$) and at the same time yields a relic density of  $\Omega_{\rm DM} h^2 < 0.12$ for the dark matter candidate $A^{(1)}_1$, such that $A^{(1)}_1$ does yield more than the observed DM. The bounds from LHC and from the dark matter relic density are complementary because a large BLKT induces a weak interaction strength thus a small annihilation cross section and a large relic abundance for a given $R^{-1}$ but the resonance search result gives a weaker bound on $(r_{B}/L, r_{W}/L)$ plane for a larger $R^{-1}$. 
Thus from LHC searches, more parameter space in the ($r_{B}/L$, $r_{W}/L$) plane is allowed for larger values $R^{-1}$ TeV, but  the allowed parameter space shrinks back with a larger value of $R^{-1}$ due to the relic density constraints. For example for $R^{-1}=2.4$~TeV (the sky blue region in Fig.~\ref{fig:relic_contour}) only small $r_W/L$ and very small $r_B/L$ are allowed by the combined bound.

{ Before coming to direct and indirect detection of DM, a comment is in order. In the determination of the relic density presented in this section, we fully considered perturbative contributions to the annihilation and co-annihilation processes. Beyond the scope of this paper, non-perturbative effect may be also important\footnote{Sommerfeld effect can lead a reduction of the relic density of up to $50 \%$ for a wino mass below $\sim 2$~TeV in a supersymmetric model~\cite{Beneke:2016ync}. The effect can be even more significant for heavier cases.} and deserves further investigation~\cite{Somm1,Somm2,Somm3,Somm4}. We would reserve the further study for the future.  }

\subsection{Direct and indirect detection of EW KK DM}

\begin{figure}[t]
  \centering
    \includegraphics[width=0.60\textwidth]{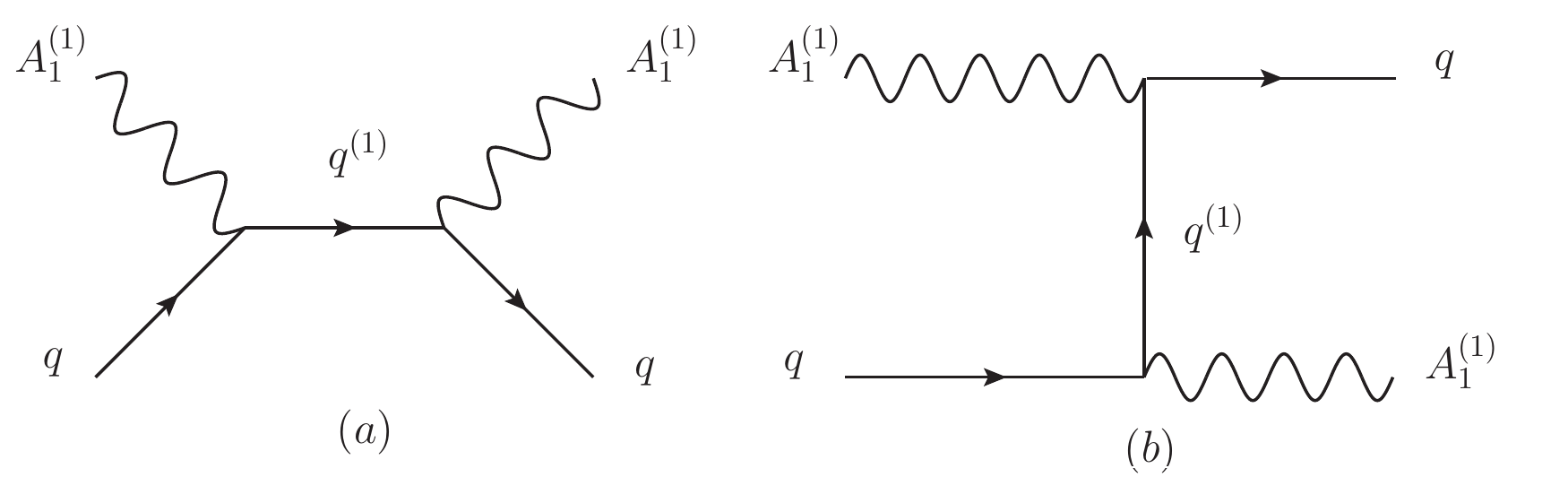}
    \includegraphics[width=0.33\textwidth]{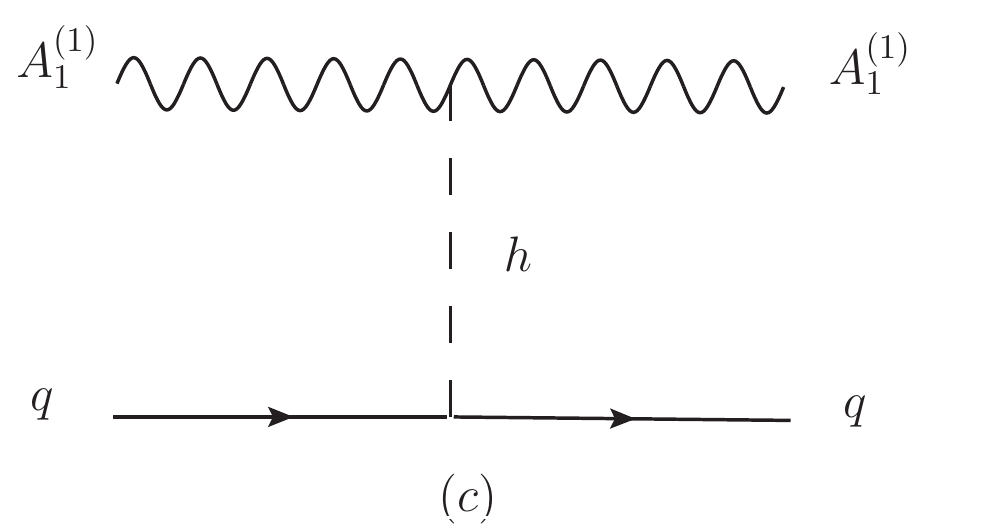}
     \caption{\label{fig:DD}
    Tree level diagrams for the elastic scattering of $A_{1}^{(1)}$ with quarks. In figures (a) and (b) scattering occurs through the level 1 KK quark and in figure (c) through the SM Higgs.}
\end{figure}

Despite many ongoing searches with DM direct detection experiments, no firm signals of dark matter have been observed yet \cite{Cushman:2013zza,Akerib:2015rjg,Tan:2016zwf}, and these experiments have set bounds on the scattering cross-section of dark matter.  
The elastic scattering of KK DM and nucleon is mediated by exchange of KK quark and the SM Higgs as shown in Fig. \ref{fig:DD}. The spin independent (SI) scattering cross-section is given as 
\begin{eqnarray}
\sigma_{SI} = \frac{M_{T}^{2}}{4 \pi (M_{A_{1}^{(1)}} + M_{T})^{2}} [Z f_{p} + (A - Z) f_{n}]^{2} ,
\label{eq:sigsi}
\end{eqnarray}
where $M_{T}$ is the target nucleus mass, $Z$ and $A$ are the atomic number and atomic mass of the target respectively. 
The elastic scattering form factor for the nucleon is given by 
\begin{equation}
f_{p/n} = \sum_{u, d, s} \frac{(\beta_{q} + \gamma_{q})}{M_{q}} M_{p/n} f_{T_{q}}^{p/n},
\label{eq:fp}
\end{equation}
where $M_{p/n}$ is the mass of the proton(neutron) and $M_{q}$ is the light quark mass. 
We adopt the nucleon matrix elements from Ref.~\cite{Backovic:2015cra} in our analysis.
The dominant contribution to the nucleon form factors is from light quarks, whereas the heavier quarks $(c, b, t)$ contribute through the gluon form factor, given as $f_{T_{G}}^{(p/n)} = 1 - \sum_{q} f_{T_{q}}^{(p/n)}$ but the effects are suppressed~\cite{Servant:2002hb}. 

The $\beta_q$ encapsulates the contributions from the left and right handed KK quarks as depicted in Fig.~\ref{fig:DD} (a) and (b) and $\gamma_q$ from the Higgs (Fig.~\ref{fig:DD} (c)):
\begin{eqnarray}
\beta_{q} &=& M_{q} (\cos \theta_{W}^{(1)} g_{B^{(1)} \psi^{(1)} \psi^{(0)}})^{2} 
\left[Y_{q_{L}}^{2} \frac{M_{A_{1}^{(1)}}^{2} + M_{q_{L}^{(1)}}^{2}}{\left(M_{q_{L}^{(1)}}^{2} -  M_{A_{1}^{(1)}}^{2} \right)^{2} } + Y_{q_{R}}^{2} \frac{M_{A_{1}^{(1)}}^{2} + M_{q_{R}^{(1)}}^{2}}{\left(M_{q_{R}^{(1)}}^{2} -  M_{A_{1}^{(1)}}^{2} \right)^{2} } \right]  \nonumber \\
 &+&  M_{q} (\sin \theta_{W}^{(1)} g_{W^{(1)} \psi^{(1)} \psi^{(0)}})^{2} \left[\frac{1}{4} \frac{M_{A_{1}^{(1)}}^{2} + M_{q_{L}^{(1)}}^{2}}{\left(M_{q_{L}^{(1)}}^{2} -  M_{A_{1}^{(1)}}^{2} \right)^{2} }  \right], \\
\gamma_{q} &=& \frac{M_{q} [(\cos \theta_{W}^{1} g_{B^{(1)}  \phi^{(1)} \phi^{(0)}})^{2} + (\sin \theta_{W}^{1}  g_{W^{(1)} \phi^{(1)} \phi^{(0)}})^{2} ] }{2 M_{h}^{2}},
\label{eq:gamma}
\end{eqnarray}
where $M_{h} \simeq125 \rm GeV$ is the SM Higgs mass. $g_{V^{(1)} \psi^{(1)} \psi^{(0)}}$ is the gauge coupling of the respective level 1 gauge boson V
with fermions as defined in Eq.~(\ref{eq:g110}), $Y_{q_{L/R}}$ are the values of the hypercharges of the SM quarks, with the convention $Y_{i} = Q_{i} - T^{3}_{i}$, $Q_{i}$ and $T^{3}_{i}$ being the electric charge and 
weak isospin respectively. $\theta_{W}^{(1)}$ is the level 1 KK Weinberg angle and $M_{q_{L/R}^{(1)}}$ is the mass of the level 1 KK quark introduced in Fig \ref{fig:DD}. The mass gap between the KK quark and dark matter masses are parameterized by 
\begin{eqnarray}
\delta_{q} 
&=& \frac{M_{q^{(1)}} - M_{A_{1}^{(1)}}}{M_{A_{1}^{(1)}}} 
\approx \frac{1-R M_{A_{1}^{(1)}}}{R M_{A_{1}^{(1)}}}, \nonumber 
\end{eqnarray}
where we used the approximate relation $M_{q^{(1)}} \approx 1/R$. 

The spin-dependent cross-section is given by 
\begin{eqnarray}
\sigma_{SD} = \frac{M_{T}^{2}}{ 6\pi \left(M_{A_{1}^{(1)}} + M_{T}\right)^{2}} J_{N} (J_{N} + 1) \bigg[\sum_{u,d,s} \alpha_{q} \lambda_{q} \bigg]^{2},
\label{eq:sigsd}
\end{eqnarray}
with $\alpha_{q}$ and $\lambda_{q}$ given as
\begin{eqnarray}
\alpha_{q} &=& 2 \left[ (\cos \theta_{W}^{(1)} g_{B^{(1)} \psi^{(1)} \psi^{(0)}})^{2} \left(\frac{Y_{q_{L}}^{2} M_{A_{1}^{(1)}}}{M_{q_{L}^{(1)}}^{2} -  M_{A_{1}^{(1)}}^{2}} + \frac{Y_{q_{R}}^{2} M_{A_{1}^{(1)}}}{M_{q_{L}^{(1)}}^{2} -  M_{A_{1}^{(1)}}^{2}} \right)  \right]   \nonumber \\
 &+&  (\sin \theta_{W}^{(1)} g_{W^{(1)} \psi^{(1)} \psi^{(0)}})^{2} \left( \frac{M_{A_{1}^{(1)}}}{M_{q_{L}^{(1)}}^{2} -  M_{A_{1}^{(1)}}^{2}} \right), 
\label{eq:alpha}
\end{eqnarray}

\begin{eqnarray}
\lambda_{q} = \Delta_{q}^{p} \frac{ \langle S_{p} \rangle}{J_{N}} + \Delta_{q}^{n} \frac{ \langle S_{n} \rangle}{J_{N}},
\end{eqnarray}
where $\Delta_{q}^{p/n}$ is the fraction of the nucleon spin carried by the quark, for which we use the values from Ref.~\cite{Backovic:2015cra}. The ratio $\langle S_{p/n} \rangle /J_{N}$ is the fraction of the total nuclear spin carried by the spin of the nucleon, $J_{N}$ being the total nuclear spin. Direct detection experiments commonly present their constraints in terms of effective WIMP-nucleon cross sections for which $\lambda_{q}$ reduces to $\Delta_{q}^{p/n}$.

\begin{figure}[t]
  \centering
      \includegraphics[width=0.6\textwidth]{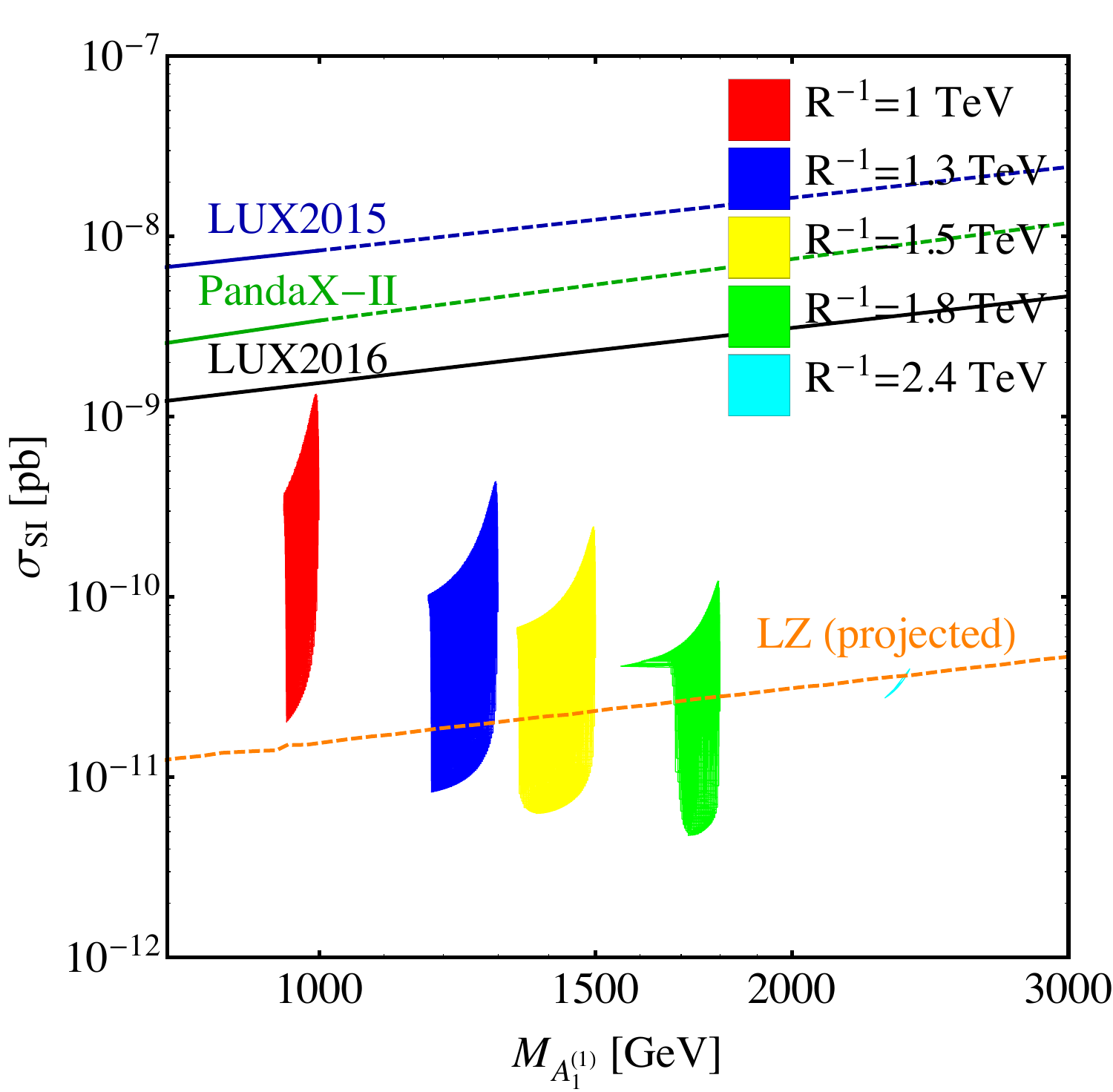}
     \caption{\label{fig:DDrwrb}
The expected SI cross sections ($\sigma_{SI}$) are plotted for various values of $R^{-1}$ ranging from 1 TeV to 2.4 TeV. For each $R^{-1}$, we vary $(r_B/L,r_W/L)$ within the allowed parameter regions from Fig.~\ref{fig:relic_contour} in order to obtain the predicted regions of $\sigma_{SI}$. Current exclusion limits are set by LUX~\cite{Akerib:2015rjg} (dark blue and black) and PandaX-II~\cite{Tan:2016zwf} (dark green). 
The expected sensitivity of LZ (orange) is found in Ref.~\cite{Cushman:2013zza}.  The LZ  projected sensitivity  would cover the entire remaining parameter space for $R^{-1}=1$ TeV.  A large part of the parameter space in 
$(r_{B}/L, r_{W}/L)$ is within the testable range for a larger value of $R^{-1}$, too.
}
\end{figure}

In Fig.~\ref{fig:DDrwrb} we show the spin independent scattering cross section within the parameter space, which is fully compatible with the currently available experimental results from EWPT, KK resonance searches and the right relic abundance of DM. The expected cross sections  are represented in the regional plots with different colors corresponding to a given value of $R^{-1}$: 1 TeV (red), 1.3 TeV (blue), 1.5 TeV (yellow), 1.8 TeV (green) and 2.4 TeV (sky blue) from left to the right. The expectations are compared with the limit on the spin independent dark matter-proton scattering cross-section from the latest LUX result~\cite{Akerib:2015rjg} and also from the PandaX-II result~\cite{Tan:2016zwf}. Current experiments are not quite sensitive enough to probe DM masses above 1 TeV. We also present the projected sensitivity limit from LUX-Zeplin (LZ), $3\times 10^{-48} {\rm cm}^2$, which is based on the estimation of a 3 year run with $6000$ kg fiducial mass~\cite{Cushman:2013zza}. It is encouraging to notice that the future LZ sensitivity region would cover the full parameter space for $R^{-1}\leq 1$ TeV and also quite large portions of the parameter spaces for heavier DM above 1.3 TeV.

\begin{figure}[t]
  \centering
      \includegraphics[width=0.99\textwidth]{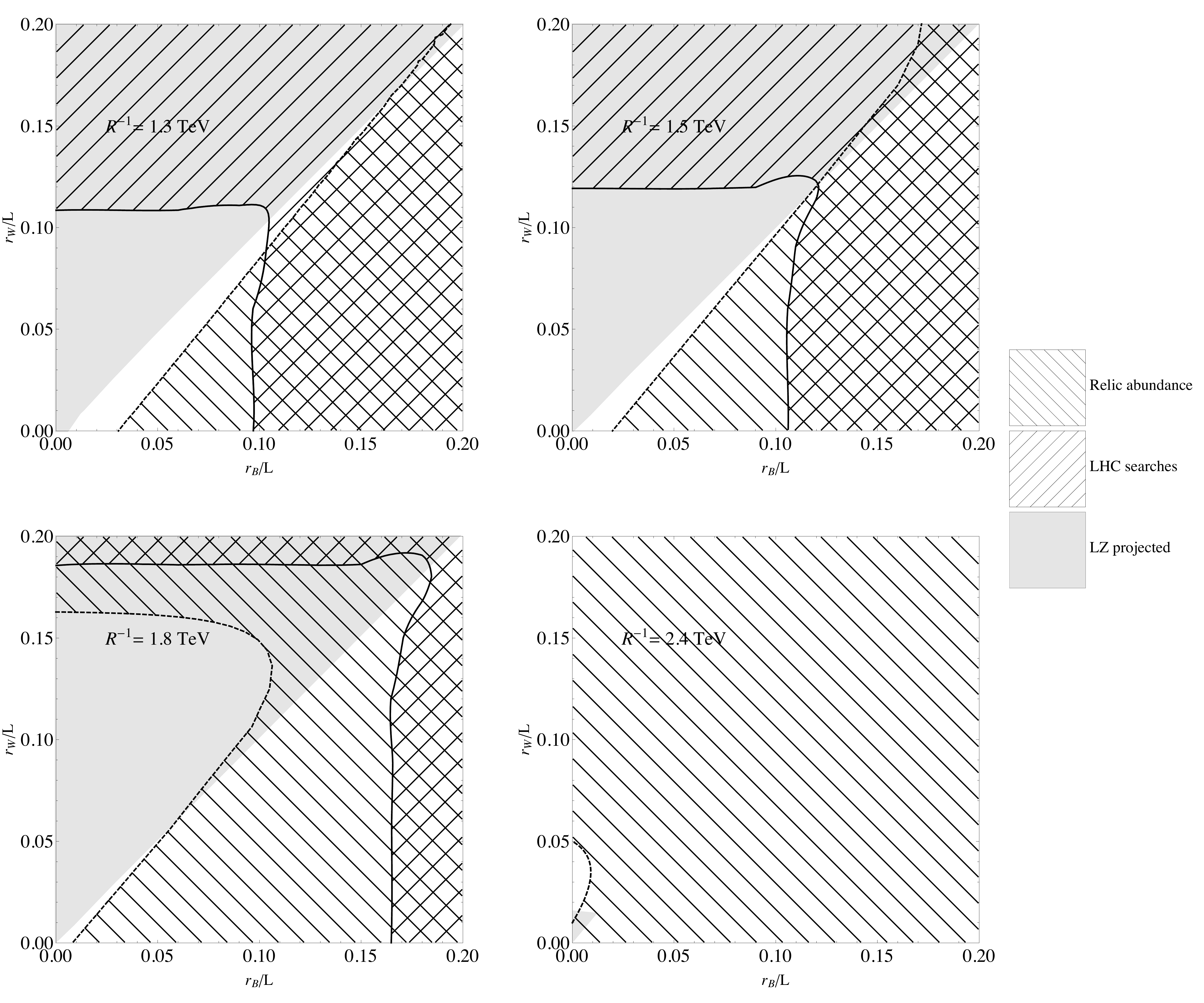}
     \caption{\label{fig:Allconstraints}
 { Combined constraints from relic abundance (left hatched), LHC collider search (right hatched) and LZ projected direct detection sensitivity (grey shaded) for $R^{-1} = 1.3, 1.5, 1.8, 2.4$ TeV in $(r_{B}/L, r_{W}/L)$ plane. The solid curves are limits from LHC collider search and the dashed curves are limits from relic abundance.} 
 }
\end{figure}

In Fig.~\ref{fig:Allconstraints}, we present the remaining parameter space after taking the complementary constraints from the collider search  and also the relic abundance calculation for $R^{-1}=1.3, 1.5, 1.8$ and $2.4$ TeV, respectively. 
 The expected coverage of the 3 year run at LZ is shaded by grey. 
It is clear that LZ can probe almost entire parameter space which is compatible with current experiments.
Thus, the DM direct detection experiments will play important complementary roles to the LHC searches and future EWPT experiments in the search of KK DM.

Finally, we comment on indirect signals of electroweak KK DM. First, the mass of KK DM (which is likely to be heavier than 1 TeV or even higher but still less than 2.4 TeV) is rather high compared to the range of energies $\lsim \text{a few}\times \mathcal{O}(100)$ GeV considered in the recent Gamma-ray studies of Dwarf galaxies and of the milky way galactic center~\cite{Hooper:2012sr}. {Other cosmic ray measurements could in principle provide constraints on heavier masses  but  these observations currently involve large astrophysical uncertainties}~\cite{Giesen:2015ufa, Hooper:2014ysa}. We have noticed several studies in this line: Refs.~\cite{Bergstrom:2004nr,Bertone:2010fn,Bonnevier:2011km,Melbeus:2011gs} study photon lines (and continuous photon background) from UED models, partially with $B^{(1)}$, partially with $Z'$ DM. However, the expected fluxes of the product particles into cosmic ray signals is highly dependent on the existence of boost factor, e.g. by Sommerfeld enhancement.
\footnote{In the heavy mass regime $m_{\rm LKP}\gsim 1000$ TeV, an enhancement of ${\cal O}(100)$ would be required to imply any bounds. See e.g. \cite{Park:2009cs}.   
} 
We reserve a dedicated study of the Sommerfeld enhancement for indirect detection in the context of UED models for the future and do not include indirect bounds into the bounds presented in Fig.~\ref{fig:Allconstraints} .

\section{Summary and discussion}
\label{sec:conclusion}

Bulk masses and boundary-localized kinetic terms in models with Universal Extra Dimensions significantly change the phenomenological properties of the Kaluza-Klein dark matter. A linear combination of KK weak boson and KK hypercharge gauge boson forms the lightest Kaluza-Klein particle (LKP), which we call electroweak Kaluza-Klein dark matter. Depending on the parameter choice, the electroweak Kaluza-Klein dark matter may be mainly KK $Z$-boson like or KK photon-like. In this paper, we perform all the detailed derivation of KK weak mixing angles with KK mass spectra and their couplings with the standard model particles taking brane localized kinetic terms for electroweak gauge bosons into account. 
We then compare our theoretical expectations with the existing experiments ranging from electroweak precision tests, LHC resonance searches as well as dark matter direct detection experiments to determine the  parameter space compatible with current observations. 
Within the setup we discussed (5D UED with BLKTs for electroweak gauge bosons) the identified upper limit on the KK dark matter mass is  extended to 2.4 TeV, which is significantly heavier than the conventionally quoted value at 1.3 TeV in minimal UED models. The heavier regime above a TeV will be tested at future experiments including LUX Zeplin (LZ) as well as at future LHC resonance searches.  Indirect dark matter searches can provide important constraints for $W^{3(1)}$-like DM, but for them to be relevant, a large boost factor is required. Sommerfeld enhancement could potentially provide such a boost factor, and it could also have a (weaker) effect on the $W^{3(1)}$ relic density, but a quantitative study of the non-perturbative effect including Sommerfeld effect is reserved for the future.

Finally, we would emphasize that the boundary terms for fermions (and their bulk masses), the Higgs, and also gluon would provide interesting collider phenomenology, which is not considered in this study. Including these terms is an interesting  task for the future.

\section*{Acknowledgments}
TF, KK and SCP would like to thank the organizers of the CERN-CKC TH Institute on Charting the Unknown: interpreting LHC data from the energy frontier, where part of this work was being completed. KK thanks Harish-Chandra Research Institute for its kind hospitality during completion of this study. 
TF's and DWK's work is supported by IBS under the project code, IBS-R018-D1. TFs work was also supported by the Basic Science Research Program through the National Research Foundation of Korea (NRF) funded by the ministry of Education, Science and Technology (No. 2013R1A1A1062597).
GM is supported by the Fermilab Graduate Student Research Program in Theoretical Physics and in part by the National Research Foundation of South Africa under Grant No. 88614 as well as the dissertation fellowship at the University of Kansas.
KK is supported in part by the US DOE Grant DE-FG02-12ER41809. SCP is supported by the National Research Foundation of Korea (NRF) grant funded by the Korean government (MSIP) (No. 2016R1A2B2016112). 

\bibliographystyle{JHEP}
\bibliography{draft}

\end{document}